\documentclass[aps,prl,groupedaddress]{revtex4}

\usepackage{amsfonts,amssymb}

\usepackage{enumitem} 
\usepackage{amsthm}
\usepackage{tikz}
\usetikzlibrary{calc}
\usetikzlibrary{arrows}
\usepackage{tikz-3dplot}
\usepackage{color} 
\usepackage[fleqn]{amsmath}
\usepackage{amsthm}
\usepackage{amssymb}
\usepackage{amsfonts}
\usepackage{bbm}
\usepackage{wrapfig}
\usepackage{epsfig}
\usepackage{times}
\usepackage{hyperref}
\usepackage{bm}
\usepackage{float} 
\usepackage{appendix} 
\usepackage{mathrsfs}
\usepackage{appendix}
\usepackage{graphicx}

\usepackage{graphicx}

\DeclareGraphicsExtensions{.png}

\newcommand{\comment}[1]{}
\newcommand{\ket}[1]{\left |  #1 \right\rangle}

\theoremstyle{plain}

\theoremstyle{definition}

\begin{document} 

\title{S-money: virtual tokens for a relativistic economy}

 \author{Adrian \surname{Kent}} \affiliation{Centre for
	Quantum Information and Foundations, DAMTP, Centre for Mathematical
	Sciences, University of Cambridge, Wilberforce Road, Cambridge, CB3
	0WA, U.K.}  \affiliation{Perimeter Institute for Theoretical
	Physics, 31 Caroline Street North, Waterloo, ON N2L 2Y5, Canada.}
	\date{\today}
	
\begin{abstract} 

We propose definitions and implementations of ``S-money'' --
virtual tokens designed for high value fast
transactions on networks with relativistic or other 
trusted signalling constraints, defined by inputs that in general
are made at many network points, some or all of which may be space-like
separated.  
We argue that one significant way of characterising types of money in
space-time is via the ``summoning'' tasks they can solve: that is, 
how flexibly the money can be propagated to a desired space-time point
in response to relevant information received at various space-time points.  
We show that S-money is more flexible than standard quantum or classical
money in the sense that it can solve deterministic summoning tasks
that they cannot.   
It requires the issuer and user to have networks of 
agents with classical data storage
and communication, but no long term quantum state storage,
and is feasible with current technology. 
User privacy can be incorporated by secure bit commitment and
zero knowledge proof protocols.   The level of privacy feasible
in given scenarios depends on efficiency and composable 
security questions that remain to be systematically addressed.  
\end{abstract}
	
		\maketitle
	
\section{Introduction}

Money, share certificates, passwords and other tokens are
familiar quantities whose definitions nonetheless
continue to evolve.   For example, our conceptions of the function and
physical form of money 
were expanded by the inventions of cryptocurrencies (see
e.g. Ref. \cite{lansky2018possible}) and Wiesner's quantum money
\cite{wiesner1983conjugate} respectively.   

Here we reconsider money and tokens from another
perspective, in which they are used on
a network of points in space-time
with a causal structure enforced by
trusted constraints.  
Our primary motivation is that relativistic 
signalling constraints imply a causal structure
that plays a significant and growing role in the
global economy.   
Since general relativistic corrections in weak 
gravitational fields are small, the background space-time around Earth
is well approximated by Minkowski space.
We take networks in Minkowski space, without any further signalling
constraints, as our main illustration; thus we write $P \prec Q$ if $P$
lies in the past light cone of $Q$, and $P \preceq Q$ if either
$P \prec Q$ or $P=Q$.  

However, our discussion applies to any fixed background space-time, 
so general relativistic corrections can also be included where
significant.  Our discussion also applies to 
network causal structures defined 
by additional constraints arising from
trusted technological limitations.  For example,
one or both parties might accept that the other cannot
practically communicate at or near light speed through the
interior of the Earth, even though physics gives several
ways to do this in principle.  For parties who cannot
communicate through the Earth's interior at speeds greater
than $\frac{2c}{\pi}$, 
the fastest signals between points on the Earth's surface
travel at light speed on a great circle.   
Parties might additionally trust lower than
light speed signalling bounds, enforced by
specific features of a given network. 
In this context $P \prec Q$ means that the parties
agree that both can signal from $P$ to $Q$.    
 
The light speed bound on communications plays
a significant role in the global financial system and
some of its implications for arbitrage are well known. 
However, the broader financial implications of 
special relativity have received
little attention, with the notably amusing exception of Krugman's 
analysis of interstellar interest rates \cite{krugman2010theory}. 

We identify two problems that are intrinsic to trades on networks with
trusted signalling constraints when transaction speed is critical. 
The first is preventing illegitimate duplication without incurring the
transaction delays required in order to cross-check across the
network.
The second is allowing the location of token presentation
to depend as flexibly as possible on incoming data across the network, so that 
the token user can effectively allow the token to ``materialise'' 
at a particular space-time point in response to relevant events 
and incoming information. 

We propose a solution in the form of ``S-money'' schemes defining
virtual tokens, of a type we describe below.    The S may stand for ``super'' in contexts 
where our schemes have potentially critical advantages over other types of money.
As we explain below, when used by parties with distributed networks
of agents in space-time, these advantages can include practicality
(compared to quantum money with current technology) and flexibility
of response to incoming data (when compared to 
all previously defined types of money, as far as we are aware).

One natural way of investigating this flexibility 
is via generalised summoning tasks.
Summoning was originally 
\cite{kent2013no, kent2012quantum} 
considered as a way of studying the properties of quantum
information in space-time.  In that context, it is defined as a task between two mistrustful parties, Alice and Bob,
who each have networks of collaborating agents. 
Bob creates a quantum state, keeping its classical description
secret, and gives it to Alice at some starting point.
Alice is required to return it at some later point that
depends on communications received from Bob at other points. 
Various versions of the task have been studied
\cite{kent2013no, kent2012quantum, hayden2016summoning, adlam2016quantum, kent2018unconstrained}.  
Here we go beyond the original context and
definitions, and think of summoning tasks more generally as 
tasks in which Alice wishes for something to become available at some specific
space-time point that depends on data that her agents may have
received at various points in space-time.   
In our case, it is a valid virtual S-money token that becomes
available for presentation at a specific space-time point on a network, which depends on
market or other data Alice's distributed agents received at various
network space-time points.   
In this context, the ``S'' may alternatively
stand for ``summonable''. 
For those unfamiliar with summoning, it is worth stressing that -- in
either context -- many more tasks are solvable than initial
intuition might suggest.  In particular, many summoning tasks
are solvable even though they cannot be solved
by making the quantum state or virtual token follow some definite
path through space-time.   

We show that S-money is more flexible than standard quantum or classical
money, in the sense that it can solve any deterministic summoning task
that they can without requiring the same resources, and many that they cannot.   
Without an extra cryptographic layer, S-money does not protect the user's
privacy: the issuer can read the user's inputs as soon as they
are provided.
However, we show that bit commitments and related cryptographic 
protocols may be used to effectively encrypt these inputs and retain user privacy. 

In summary, our aim here is to introduce and explore the conceptual framework 
of S-money and show that it has some significant advantages in
specific contexts.
Other types of money also have major advantages, which
are not replicated by the types of S-money described here.   
It may well be possible to improve S-money further and/or
to devise hybrid schemes that combine
some or all of the advantages of S-money and of other types.
We hope researchers will begin addressing these questions;
we discuss some of the relevant issues later.   
  
The two problems S-money addresses are not new in principle.
Indeed, in one sense they were larger in past,   
given the low technological bounds on signalling speeds before
the development of the telegraph and radio.\footnote{These certainly slowed the responsivity 
of the global financial system and led to significant arbitrage opportunities.
For example, it has been suggested that Nathan Rothschild was able to 
profitably exploit early information about the outcome of 
the Battle of Waterloo.   It should be noted, though, that
historical evidence conflicts with the popular anecdotes, which may largely be inventions by 
adversaries of the Rothschilds: some discussion can
be found in Ref. \cite{ferguson1999house}.  The anecdotes
nonetheless at least show that the existence of signalling arbitrage opportunities was widely
recognised in the 19th century.   It would be very surprising if such opportunities
were not sometimes exploited.}
The relatively undeveloped state of the pre-1830s global financial
system, computing technology and cryptography presumably help to explain
the lack of historical precedents.

\section{Networks and agents}

We are interested in token schemes that allow some form of access 
-- for example to goods, services, data, physical objects or environments.       
Depending on the context, such tokens might for example play the role
of money, passwords, keys or passcards.
They may be used by parties with many possible relationships.
The parties may be individuals or may be networks of 
collaborating agents who trust one another.  

For illustration, we suppose 
that the scheme involves money issued by one agency, $B$ (Bob, the
Bank, or issuer) to another $A$ (Alice, the Acquirer, or user).
We suppose that $A$ and $B$ have pre-equipped 
themselves to carry out communications and transactions
at a pre-agreed finite set of space-time points defining a network. 

A network point $P$ here represents a small pre-agreed
local space-time region around the point $P$, which can 
contain separate secure sites for user and issuer agents,
and authenticated classical and/or quantum channels
between these sites.  All communications associated with
a given network point must take place within the associated
space-time region.   Typically, when there is a natural 
fixed frame choice to describe the network, we assume that the spatial
diameter of these regions is significantly smaller than the
spatial separation between any pair of network points. 

Unless otherwise stated, we assume below that the only signalling constraint
is the impossibility of superluminal communication, and that
both $A$ and $B$ can send secure classical signals (for instance
using one-time pads maintained by quantum key distribution) from any network point
to any other network point within its future light cone.    

As usual in relativistic cryptography \cite{kentrelfinite},
we assume that each agency may be represented by a set of agents
who are completely trusted and whose actions are coordinated, so 
that they can be identified as effectively a single party. 
The parties may have unequal technological
resources. 
We suppose that $B$ plays the role of a universal bank, and has an agent 
available at each network point; the same agent may be available at 
many timelike separated points. 
$A$ may be a single user or small organisation, with only one or a few
agents, moving between network points, perhaps
in response to locally acquired data.   Alternatively, she may
be an organisation similar to $B$, with agents available at every
network point. 

\subsection{Authentication}

We assume that at each network site user and issuer agents are
able to authenticate each other's communications.
This may be by physical means or by using some standard
cryptographic authentication scheme, or a combination. 
The authentication is of the issuer and user rather than
of individual agents.   Each issuer agent is able to verify that a 
communication came from the user, and vice versa.
They do not necessarily need to be able to verify that 
the communication came from a specific agent of the other
party, nor do they need to authenticate the location of that party. 
It is sufficient for our schemes that a party receives
authenticated communications from the other at particular space-time
points.  

We also assume that all user agents have secure authenticated communication
channels with one another and that the same applies to all issuer
agents.   Since we assume perfect trust among user agents, the
user agents may simply identify themselves and their locations 
in these communications, if required; similarly for communications
among issuer agents.   

\section{The problem of duplication}  

\subsection{Classical data money in space-time}

Consider first a simple money scheme based on classical data strings.
(Essentially the same scheme, though without the relativistic network
context, was discussed by Gavinsky \cite{gavinsky2012quantum}.)
To issue a money token, $B$ generates a secure random string $S$ at
some point $P_0$ and transmits it securely to $A$ at some network
point $P \succeq P_0$.  
The string has an agreed monetary value $V(S)$.  
$B$'s agent at some point $P_1$, where $P_0 \preceq
P_1  \preceq P$, securely communicates the values $(S, V(S))$
to all $B$'s agents in the future light cone of $P$.   
$A$'s agent at $P$ may securely carry the string $S$ to some
network point $Q \succ P$, or securely transmit it to another agent
of $A$ at such a point.    Here we write $Q \succ P$ if and only if
$Q$ is in the causal future of $P$.   After some sequence of such actions, $A$'s 
agent at a network point $Z \succ \ldots \succ Q \succ P$ may present 
the string to $B$'s agent at $Z$, who checks against his records,
validates it, and issues resources to the value $V(S)$.     
Having done so, he communicates the fact to all $B$'s agents
in the future light cone of $Z$, ensuring that they will
not accept the string if it is re-presented to them.  

Among other problems, this scheme is vulnerable to illegitimate 
multiple presentations of the same token.   
Once $A$ receives the string $S$, she may copy
it and transmit it to agents at two or more pairwise space-like
separated network points $Z, Z', \ldots$, who may present it to 
$B$'s agents at each of these points.   None of $B$'s agents
at these points 
can be aware that the string is also being presented at the other
points.  If they follow the scheme as defined, they thus 
each issue resources to the value $V(S)$, allowing $A$
illegitimate access to multiple resources.

\subsection{Detection of fraud}

$B$'s agents will communicate their acceptance to one
another, so that $B$ will eventually become aware of $A$'s fraud.
Legal and/or other sanctions may follow.   
These may be a sufficient deterrent in some scenarios, 
but not all.   Fraudulent transactions may be difficult
to reverse, particularly if $A$'s agents exploit, consume, abscond
with or trade on resources.  The legal situation may be complex; each
agent of $A$ may claim that they behaved honestly, unaware
of the others' actions.   If $A$'s motive is, say, sabotage of 
a financial system, she may not be overly concerned about detection.     

\subsection{Cross-checking and delays} 

For these reasons, among others, $B$ may want to make the scheme
secure against multiple presentation.    One possible security mechanism is for any agent of $B$
who receives the money token to cross-check with all other agents of $B$
(either individually or via a central server) and make sure they
are aware he has received the token, and have not themselves
previously received it, before he issues the resources. 
If access delays of the order of twice the network diameter are
acceptable, this may be a viable solution.   
For global markets with superfast local trading responding to local
data, let alone future interplanetary and 
interstellar markets, it will not be.  

\section{Preventing duplication without cross-checking}

We now consider a simple
alternative proposal, fixed path classical data money. 
This can be thought of as a special case of the S-money
schemes we define below.  
Like general S-money schemes, it prevents illegitimate 
duplication.   However it lacks the flexibility
of general S-money schemes.  

\subsection{Fixed Path Classical Data Money} 

We define {\em fixed path   
classical data money} to be classical data money
with an extra condition on the user, $A$, that 
ensures that the money token follows a single 
path through the network. 
When $A$'s agent at 
$P$ receives the string $S$, she is required to tell
$B$'s agent there at which point $Q\succ P$ she wants the token
to be valid.   $B$'s agent at $P$ communicates this to
$B$'s agent at $Q$, who will only accept the token as valid
if he has received this confirmation.  

If $A$'s agent at $Q$ wants to propagate the money token
further, then she does not present the string to $B$'s 
agent there, but tells him at which point $R\succ Q$ she wants
it to be valid.   
$B$'s agent at $Q$ communicates this to
$B$'s agent at $R$, along with confirmation that 
he received a communication from the agent at $P$ confirming
that the token propagated from $P$ to $Q$. 
The agent at $R$ will only accept the token as valid
if he has received these confirmations.  

This continues until the token is
presented by $A$'s agent at $Z$ to $B$'s agent at $Z$.
He  checks that $S$ is
the correct string and that $B$'s agents at $P, Q, R \ldots$ 
have confirmed that the token was validly propagated from $P \rightarrow Q$,
$Q \rightarrow R$, and so on.   $A$ thus cannot validate the
token at two spacelike separated points $Z, Z'$, since there
cannot be two distinct valid propagation paths $P \rightarrow \ldots \rightarrow Z$
and  $P \rightarrow \ldots \rightarrow Z'$.

\begin{figure}[h]
\centering
\includegraphics[width=0.5\textwidth, height=5cm]{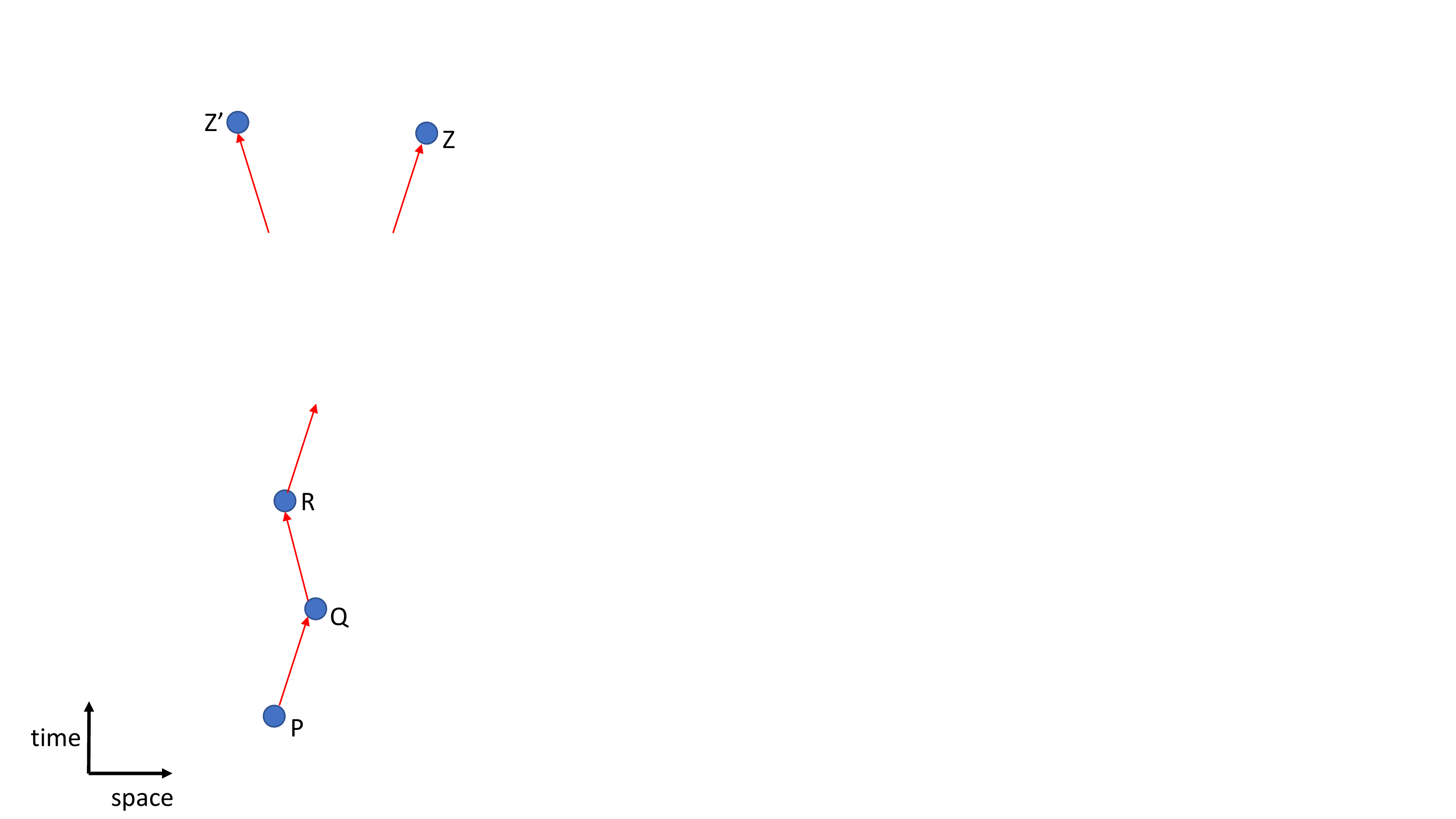} 
\caption{Schematic description of fixed path classical data money.}
\label{one}
\end{figure}

\subsection{Drawbacks of Fixed Path Classical Data Money}

Clearly, in this scheme, $B$ knows
ahead of time where the money token will be valid, 
so that the user's privacy is compromised.   
This can be addressed.   We describe later 
ways in which fixed path classical data money, and 
the more general S-money schemes we consider below,
can be extended to ensure $A$'s privacy below. 

Another concern may be that the scheme reduces $A$'s flexibility.
In a general network, $A$'s agent at $P$ may not know at 
that point which site the token should go to next, and may 
wish to decide later.   In principle, this concern can be 
significantly mitigated by refining the network to include more decision
points.    If the token were a physical object
(which may be massless and travel at light speed) being
propagated by $A$, then at any given point $P$ the user $A$ would 
know the token's current planned trajectory and hence,
to good approximation, a near future space-time point $Q$ it will
reach.   A sufficiently refined network would allow her
to propagate the token from $P$ to $Q$, reevaluate the trajectory
there, and so continue.  
If one thinks of the user as an individual moving about the
planet carrying a physical token, for example, then in principle
our scheme can emulate this if the user carries 
a mobile device exchanging signals with issuer's local stations, 
which can calculate her velocity at any given point and propagate the token to an
appropriate future network point.\footnote{We suppose here that the
network of local stations is sufficiently dense for the applications
envisaged.}
Similarly, our scheme can emulate a data token sent at light speed over 
a user's network, by taking switching points to be points of the 
scheme's network and propagating our token between switching points
according to the scheme.  

A less obvious drawback is that, as the name implies,
fixed path classical data money tokens follow a definite path through space-time. 
We now argue that relativistic or other signalling constraints 
motivate a more general concept of money token, which does not require 
definite token paths. 

\section{Money as a solution to tasks in space-time }

\subsection{Money and information}  

We start by proposing a way of thinking 
about money and other tokens in general, whether or not signalling
constraints apply.   This is to consider money and tokens as solutions to 
problems that arise because the information users receive over 
time is significant for their decisions and generally unpredictable
in advance.   
Money and tokens allow users (qua consumers) to acquire or access resources,
and (qua producers) to sell or 
allow access to resources, 
when they receive information that tells them this is possible 
and advantageous.   If we had perfect advance information, we
could pre-arrange every transaction that we will wish to carry
out during every trip, and leave
our wallets at home.   We could even pre-arrange a set of
contracts at birth,
committing ourselves to a lifetime of balanced production and 
consumption, signing off on the payment for the choir at our
funeral before exploring our crib.  
In fact, we don't know how we or the world will evolve, 
and so we don't know what we will want, nor what the world
will want from or offer us.  So we generally carry
some form of money, in order to buy or sell when relevant
information arrives. 
  
Relativistic and other signalling constraints impose limits on
when and where specific information can arrive.   For everyday
decisions, relativistic constraints are currently rarely relevant.
Lemon prices in Australia rarely fluctuate fast enough to 
affect whether a consumer in the U.K. should
buy lemonade, however precise her decision calculus
and however avidly she tracks the global lemon market.   
For high-frequency financial market trades
and other trades involving relativistic arbitrage, though,
they are crucial.    

Even when relevant, relativistic constraints do not 
fundamentally change the role of money for individuals,
so long as they act only as individuals.   
If someone follows a definite path in space-time, 
and makes all their financial decisions themselves, during
that path, their decisions at any point on the path are 
made only on the basis of information that has reached them
up to that point on the path.    This is true in
either Galilean or Minkowski space-time. 

However, this no longer holds true in Minkowski space-time when individuals act as agents 
of a larger enterprise.  A team of separated collaborating agents
each acquire information on their own paths.
The team thus may receive information at various points distributed throughout 
space-time, which may be shared as fast as signalling constraints
allow.  This shared information may make a persuasive case
for buying or selling at various other points in space-time.    
For money to enable buying or selling, it needs to be available
at the relevant points.  So a money token scheme needs to solve
the problem of responding to information distributed throughout
space-time, by making money available at appropriate points,
while preventing illegitimate duplication.    

A team of user agents who have only one money token, which 
they propagate along a definite causal path, only
has the option of using it at points on that path.   
The present state of such a token is defined at any
given time in an agreed frame (or more generally on any given hypersurface)
by its location.  The evolution of its future state is determined by
its velocity and higher derivatives.  These can only be 
affected by information in the past light cone of its present location.  
For this reason, fixed path classical data money and other
schemes that require money tokens to follow definite paths in
space-time do not allow optimally flexible responses to incoming
information
distributed over space-time, as the examples below illustrate. 

\subsection{A model environment for money token schemes}

We define a model environment for money tokens in which
local user agents may acquire new information 
at specific points in space-time and agents collaborate
on decisions to acquire or access resources.  
We will assume here that the relevant information and resources are classical.\footnote{Users might also
acquire quantum information, which might be delocalized.   
They might also want to acquire delocalized quantum resources,
such as singlets shared between specified sites.   Our token schemes
can allow for the possibility that users may acquire quantum information, if users generate classical
token inputs from acquired quantum information.   Our schemes could also be generalized to
allow acquired quantum information to be token inputs,
although the propagation of such data is obviously constrained 
by the no-cloning and no-broadcasting theorems.   
Our generalized schemes
involving subtokens may be used to allow users to acquire delocalized
quantum resources.  In such scenarios,
the onus is on the user to ensure that their requests are appropriately
coordinated as well as valid.   For example, if the user requests
a component of singlet $s$ at point $P$ and a component of singlet
$s' \neq s$ at a spacelike separated point $Q$, she will not acquire
a singlet shared between $P$ and $Q$, even if each request is 
valid according to the scheme.}
User agents at some set of {\it input points} $\{ P_i \}_{i=1}^m $ receive
incoming classical information, which may for example be
integers $n_i$, real numbers $x_i$ or real vectors $\underline{x}_i$.  
We also allow the possibility that the user agents at some or all of the
$P_i$ receive no information.

The user agents collaborate in order to optimize their decisions about
acquiring a local resource at some {\it presentation point} $Q$.   
There is a finite set  $\{ Q_i
\}_{i=1}^n$ of allowed presentation points. 
The user agents' collective decisions may either select one of the $Q_i$ or 
select none of them, in which case they effectively choose not to 
acquire the resource.   
We suppose that the sets $\{P_i \}$ and $\{Q_i \}$ are known
well in advance.   They may overlap: the user agents at some
or all of the $Q_i$ may receive incoming classical information
relevant to their choice of $Q$.    

For example, the global financial network defines market prices
at many different market locations at any given time, and these 
may be available to local user agents in real time.   
These prices in turn are determined in response to locally acquired
information, which may for example include revealed actions of market participants
or independent actors, physical events, and the outcomes of local
computations.   Users may wish to use information from many points
in space-time to inform their decisions about acquiring a local resource at
some given point $Q$.   

\section{Defining S-money}

S-money schemes define virtual tokens that allow users to solve
generalised summoning tasks defined by incoming information distributed over a space-time
network.   In schemes where the user has a free choice of the valid
presentation point, these tokens are generated, according to pre-agreed rules,
by inputs from the user at various network points, generally including
space-like separated points.   In schemes where the user and issuer
jointly determine the valid presentation point, the tokens are
generated by such inputs from both parties.  

User privacy is not essential to solve such tasks, and 
so we may define unencrypted S-money schemes, in which the 
user keeps nothing private from the issuer.   In these schemes, 
the issuer's agents learn the user's agents inputs immediately, 
and the issuer may be able to predict in advance where the virtual
token will be presented.    We later discuss versions of
the schemes that use cryptography to preserve user privacy.   

The definitions of S-money schemes given below are quite general. 
For example, unless we insist on instant verification by both
user and issuer as part of the definition rather than a desirable
feature attainable in many interesting examples, they include schemes that require cross-checking
by the issuer of the type described above.   
If we were to insist on instant verification by both user and issuer
as part of the definition, then we would exclude potentially interesting S-money
schemes that may require some short delay for checks without
requiring full cross-checking across the network.  
We prefer to leave the definitions general (so that some S-money 
schemes have no evident advantage)
and illustrate their scope with examples where advantages are evident.

\subsection{S-money with free user choice of the presentation
  point} 

We are interested in schemes that allow virtual tokens to be generated
and presented with some or all of the following properties:
\begin{enumerate}[label=(\roman*)]
\item ({\it free user choice}) the user may  
present a valid token at one of a number of pre-agreed space-time
points, some or all of which may be space-like separated.   
The identity of the point at which the token is valid depends solely
on data that the user's agents input into the token scheme.  
\item ({\it instant verification by issuer}) the issuer may
verify at any such pre-agreed point $Q$ whether the token is valid at $Q$. 
\item ({\it instant verification by user}) the user may 
verify at any such pre-agreed point $Q$ whether the token is valid at $Q$. 
\item ({\it security against duplication}) the scheme guarantees to the issuer 
that valid tokens cannot be presented at two or more points. 
\end{enumerate}
Of these, (iv) is crucial.  
A valid token must eventually be verified by the issuer, even if the
scheme does not satisfy (ii), and so multiple valid tokens implies
multiple spending.  
Properties (ii) and (iii) can be relaxed somewhat -- for example
by allowing verification delays that are short compared to the network
diameter -- while still allowing schemes that are advantageous
compared to cross-checking schemes.
The cases where property (i) holds or not are independently
interesting, as we discuss below.  
The various schemes we describe below all satisfy (ii)-(iv).   

We now give general definitions of {\it S-money token schemes}.
They can be used on a very broad class
of networks. In general they define tokens that are not required to follow
definite paths in space-time.   

The first definition below describes the simplest versions of these
schemes,
which have free user choice, instant
verification by issuer and user and 
security against duplication.
It is illustrated in our discussions below of classical
emulation of summoning quantum money and of Examples $1$ and $2$.  
\vskip 10pt
{\bf S-money (with free user choice and optimal issuer and user communication)} \qquad 

\begin{enumerate}[label=(\roman*)]

\item \label{first} user agents at each network point $P_i$ in an agreed set $\{ P_1 , \ldots
, P_n \}$ communicate to the corresponding issuer agent at $P_i$
classical information encoding statements 
of some pre-agreed form.\footnote{Here and in later definitions, this includes the 
possbility of the null statement, corresponding to no communication.}
These statements collectively
constrain the network points $Q$ at which the token may potentially
be valid. 

\item \label{second} The issuer agents at each $P_i$ communicate
the statements they received to all issuer agents at points
$Q\succ P_i$.    

\item \label{third} The user agents at each $P_i$ communicate
the statements they transmitted to all user agents at points
$Q\succ P_i$.    

\item \label{fourth} The token is valid at a network point $Q$, and
  should be accepted by the issuer's 
agent at $Q$ 
if presented there, 
if and only if the statements communicated to the issuer and user
agents at $Q$ imply
that (i) the token is potentially valid at $Q$, (ii) the statements
received by issuer and user agents at any
network point $R$ spacelike separated from $Q$ exclude  
the possibility that the token is also potentially valid at $R$.

\item \label{fifth} if a valid token is presented and accepted at $Q$,
$B$'s agent there communicates this to all $B$'s agents at
points $Q' \succ Q$, so it will not be accepted again at any such
$Q'$.  Similarly $A$'s agent communicates this to all $A$'s agents
at point $Q' \succ Q$, so that they are aware that it will not be accepted again at any
such $Q'$. 
\end{enumerate}

Note that conditions \ref{first}, \ref{second},
\ref{third} and \ref{fifth} all describe communications
that are relevant in applying condition \ref{fourth}. 

Conditions \ref{first} and \ref{fourth} give separate definitions of 
potentially valid and actually valid tokens, which are convenient
in understanding some of the examples we discuss.  
These definitions can alternatively be combined, so that a token
is valid at $Q$ if the set of statements received at $Q$ belong
to a list of valid sets.   In this case, every valid set of 
statements must have the property that they exclude the 
possibility of a valid
set of statements being received at any point $R$ spacelike separated
from $Q$.   The analogous definitions for the other types of 
S-money discussed below can similarly be combined.  

Conditions \ref{second}, \ref{third} and \ref{fifth} ensure that each
issuer (respectively user) 
agent communicates all relevant information to all issuer
(respectively user) agents
in his (her) causal future.
These are useful but not essential simplifying
features.  We may instead allow 
the issuer and user agents to follow agreed weaker
rules, with the tradeoffs that the criteria for token
validity are more stringent and that instant user
verification may not be possible, as follows. 
\vskip 10pt
{\bf S-money (with free user choice, without optimal issuer and user communication):}\qquad 
\begin{enumerate}[label=(\roman*)]
\item \label{prime:first} user agents at each network point $P_i$ in an agreed set $\{ P_1 , \ldots
, P_n \}$ communicate to the corresponding issuer agent at $P_i$
classical information encoding statements
of some pre-agreed form.
These statements collectively
constrain the network points $Q$ at which the token may potentially
be valid. 

\item \label{prime:second} The issuer agents at each $P_i$ communicate
the statements they receive to some subset $S_i$ of the issuer agents at points
$Q\succ P_i$.   The subsets $S_i$ are known to both parties in advance
of the protocol.   

\item \label{prime:third} The user agents at each $P_i$ communicate the statements they
transmitted to some subset $T_i$ of the user agents at points
$Q \succ P_i$.  

\item \label{prime:fourth} The token is valid, and should be accepted by the issuer's agent
at a network point $Q$ if presented there,  
if and only if the statements communicated to $Q$ imply
that (i) the token is potentially valid at $Q$, (ii) 
the statements received by the issuer agent at every
network point $R \neq Q$ exclude  
the possibility that the token is also potentially valid at $R$.

\item \label{prime:fifth} If a valid token is presented and accepted at $Q$,
$B$'s agent there communicates this to some subset $S_Q$ of $B$'s agents at
points $Q' \succ Q$.  The subsets $S_Q$ are known to both parties in
advance of the protocol.  Similarly $A$'s agent there communicates this to
some subset $T_Q$ of $A$'s agents at points $Q' \succ Q$.

\end{enumerate}

Note that conditions \ref{prime:first}, \ref{prime:second},
\ref{prime:third} and \ref{prime:fifth} all describe communications
that are relevant in applying condition \ref{prime:fourth}. 

The subsets in condition \ref{prime:second} and \ref{prime:fifth} are allowed
to be the empty set or the full set of all points in the causal future
of $P_i$ and $Q$ respectively.   Of course, if every subset is the 
full set of all points in the causal future of the relevant point,
this becomes a S-money scheme with optimal issuer and user communication.
Schemes without optimal issuer and user communication are defined to
allow instant verification by the issuer.   
A sufficient condition to
ensure instant verification by the user is that the subsets satisfy $S_i \subseteq T_i$ for
all $i$ and $S_Q \subseteq T_Q$ for all $Q$.  
In schemes without instant verification by the user, it may
sometimes be the case that the issuer would accept the token
as valid at some presentation points, but the user does not
know this at the relevant points.

For simplicity, we focus on schemes with optimal issuer and user communication in the 
examples below. 

\subsection{S-money with valid presentation points jointly
  determined by user and issuer} 

We are also interested in S-money token schemes in which both the user and issuer 
input data into the scheme, such that the identity of the point at 
which the token is valid is a joint function of user and issuer
inputs.   By definition, such schemes do not have free user choice. 
Examples $3$ and $4$ illustrate such schemes, in which the user and
issuer jointly determine the point at which the token is valid. 
We give a general definition of these schemes, 
assuming optimal issuer and user communication.

\vskip 10pt
{\bf S-money (without free user choice, with optimal issuer and user communication)} \qquad 

\begin{enumerate}[label=(\roman*)]

\item \label{afirst} user agents at each network point $P_i$ in an agreed set $\{ P_1 , \ldots
, P_n \}$ communicate to the corresponding issuer agent at $P_i$
classical information encoding statements
of some pre-agreed form.
Issuer agents at each $P_i$ also communicate to the corresponding user
agent at $P_i$ classical information encoding statements
of some pre-agreed form.
The issuer may be required to respect pre-agreed constraints relating their
statements.\footnote{For example, they may be required to make the
same statement at each $P_i$, as in Example $3$ below.}
These statements collectively
constrain the network points $Q$ at which the token may potentially 
be valid.   

\item  \label{asecond} The issuer agents at each $P_i$ communicate
the statements 
they sent and received to all issuer agents at points
$Q\succ P_i$.    The user agents at each $P_i$ likewise 
communicate the statements they sent and received to all user agents at points
$Q\succ P_i$. 

\item  \label{athird}
The token is valid at a network point $Q$, and should be accepted by the issuer's agent
at $Q$ if presented there, 
if and only if the statements communicated to the issuer and user
agents at $Q$ imply
that (i) the token is potentially valid at $Q$, (ii) the statements
received by issuer and user agents at any
network point $R$ spacelike separated from $Q$ exclude  
the possibility that the token is also potentially valid at $R$, 
assuming that the issuer has respected the pre-agreed constraints
relating their statements. 

\item \label{afourth} If a valid token is presented  and accepted at $Q$,
$B$'s agent there communicates this to all $B$'s agents at
points $Q'\succ Q$, so it will not be accepted again at any such $Q'$.
Similarly, $A$'s agent there communicates this to all $A$'s agents
at points $Q' \succ Q$. 
\end{enumerate}

Note that conditions \ref{afirst}, \ref{asecond},
and \ref{afourth} all describe communications
that are relevant in applying condition \ref{athird}. 

As before, the assumption of optimal
communications is made to simplify the
scheme, but is not essential.   
We may instead allow 
the issuer and/or user agents to follow agreed weaker
rules, with the tradeoffs that the criteria for token
validity are more stringent and that instant user verification
may not be possible.   
The definition of schemes without optimal user and issuer
communication parallels that given above for 
schemes with free user choice.  

The definition allows the possibility of pre-agreed constraints
relating the issuer's inputs, which a dishonest issuer may violate,
but does not allow a similar possibility for the user.
This makes sense in asymmetric scenarios in which the 
reputational costs to a cheating issuer are higher than
those to a cheating user, and the potential consequences of 
user cheating are more serious.    
Of course, if users perceive the risks or costs
of issuer cheating as too high, they may prefer
schemes in which the issuer's inputs are not 
required to be related.    We discuss these
points further in Examples $3$ and $4$ below.

\section{Example: Emulating quantum money with S-money} 

\subsection{ Advantages of quantum money in relativistic scenarios}

Wiesner's quantum money \cite{wiesner1983conjugate} uses quantum information
embedded within a physical token as a security mechanism.
One well understood property of quantum money is its effective unforgeability
\cite{wiesner1983conjugate, molina2012optimal}.
This unforgeability also holds for other versions of quantum 
money \cite{pastawski2012unforgeable,gavinsky2012quantum,aaronson2012quantum} that 
require the user to hold specific quantum information locally in 
order to allow their quantum-enabled token to be validated.  

What has perhaps not been sufficiently emphasized to date is that
the advantages of using quantum states rather than classical states
for money tokens are much clearer in a relativistic context than in
a non-relativistic context, for at least two reasons.   

1.   {\em The advantage given by the unforgeability of quantum tokens is 
effectively reproducible in principle by cross-checking in non-relativistic contexts,
but not in relativistic contexts.}

Unforgeability implies that a quantum token cannot be presented at two
space-like separated points, and this gives an advantage that cannot
be replicated using standard classical tokens. 

In contrast, in effectively non-relativistic scenarios, in which communication
times are negligible, effectively instantaneous cross-checking 
gives another theoretical solution to the problem of duplication.    
Admittedly, cross-checking requires  
potentially large communication and data
storage resources.   However, quantum memory technology may 
also require costly resources, if and when such technology
becomes available.    It seems plausible that, 
in many or even possibly all scenarios, classical data money 
with cross-checking may always be cheaper than using quantum money.  
If so, unforgeability may be
a good reason for preferring quantum money mainly or even only in
scenarios where relativistic (or other) signalling constraints are 
relevant and cross-checking delays are a significant 
issue.

2.   {\em Quantum tokens may be delocalized in space-time and used to solve
summoning and other intrinsically relativistic tasks that 
standard classical tokens cannot.}

Unknown quantum states, such as those stored in quantum money tokens,
can be delocalized and effectively 
propagated along multiple paths via teleportation and secret
sharing.  This allows users to respond to distributed data
in ways that are impossible if they use standard classical tokens
that must follow a fixed path in space-time 
\cite{kent2012quantum,hayden2016summoning,
adlam2016quantum, kent2018unconstrained}.   We discuss this further below. 

\subsection{Quantum money and summoning}

\subsubsection{Summoning}

Summoning was originally 
\cite{kent2013no, kent2012quantum} 
defined as a task between two mistrustful parties, Alice and Bob,
who each have networks of collaborating agents. 
Bob creates a quantum state, keeping its classical description
secret, and gives it to Alice at some starting point.
Alice is required to return it at some later point that
depends on communications received from Bob at other points. 
Various versions of the task have been studied
\cite{kent2013no, kent2012quantum, hayden2016summoning, adlam2016quantum, kent2018unconstrained}.  

We will consider here a form of summoning described
in Ref. \cite{kent2018unconstrained}. 
Alice receives 
an unknown quantum state at the start point $P$.  
Alice must either produce the state at some presentation point
$Q \succ P$, where $Q \in \{ Q_1 , \ldots , Q_N \}$ or,
if there is no valid presentation point, she should not 
return the state anywhere.\footnote{In Ref. \cite{kent2018unconstrained}
and elsewhere in the summoning literature, the $Q_i$ are referred
to as return points.  We use the term presentation point here 
so that the same term can be used in the case of quantum
money tokens, emulations of quantum money tokens by
S-money, and tokens in general S-money schemes.
We prefer to refer to S-money tokens being presented
rather than returned, since the latter term suggests 
restoring some classical or quantum physical object to
its originator.}
We write $Q = \emptyset$ in 
the latter case. 
In our examples we will assume that there is at most one
valid presentation point, although our discussion also applies to the 
case where there may be any number of valid presentation points if there
is a quantum algorithm that always chooses the same valid presentation point
for a given set of inputs \cite{kent2018unconstrained}.
So we have $Q \in \{ \emptyset , Q_1 , \ldots ,
Q_N \}$.  
The identity of $Q$ depends on 
classical information received at some set of input points $\{ P_i
\}_{i=1}^M$.   The point $P$ and the sets $\{ P_i \}$ and $\{ Q_i \}$ are known to both
parties in advance of the
task.  The sets may overlap, and may include $P$.
We assume that $Q_i \succeq P$ for all $i$.  

We take the sets of input and presentation points to be finite. 
The classical information Alice receives at $P_i$ is an 
integer $m_i$ in the range $0 \leq m_i \leq ( n_i - 1 )$. 
Thus if $Q(
m_1, \ldots , m_M )  \in \{ Q_1 , \ldots , Q_N \}$
then Alice should return the state to $Q ( m_1 , \ldots , m_M )$, 
and if $ Q( m_1, \ldots , m_M )= \emptyset$ then she should not
return the state.   
Alice knows the functional dependence of $Q$ and the 
set  $\{ n_i \}_{i=1}^M$ in
advance of the task; she learns the values of the $m_i$ only
at the points $P_i$.   

Note that we do not assume that $Q_i \succeq P_j$ for all $i,j$; 
the form of the function $Q ( m_1 , \ldots , m_N )$ may allow
an agent at some or all of the $\{ Q_i \}$ to know whether or
not their $Q_i = Q ( m_1 , \ldots , m_N )$ even though they
do not know all the inputs $m_j$.

\subsubsection{Scenarios for summoning quantum money}

Quantum money schemes, in their simplest form, also involve
two parties, who may have networks of collaborating agents,
and who play the roles of issuer (Bob, the Bank)
and user (Alice, the Acquirer), as in the classical money schemes
discussed above.   As in the case of summoning, Bob creates a 
quantum state, keeping its classical description secret, and gives
it to Alice at some starting point, in this case as part of a 
quantum money token.  

A natural scenario which combines summoning and quantum money 
is that Alice is the acquirer of a quantum money token and 
wishes to get its quantum state to some point $Q$
in space-time, in order to spend the quantum money there.
The point $Q$ -- the best place to spend the money -- is determined by classical information she receives
at other points in space-time.    In this case we can identify
the parties called Alice in the definitions of summoning and of quantum
money: there is one party, Alice, who has quantum money and needs to satisfy a summoning task 
in order to get its quantum information to the right point.   

In some scenarios the parties called Bob might also be identified as
a single party: the issuer of the quantum money might, for some
reason, be supplying information that effectively guides Alice as to where to redeem it.  In general, though, it
is more natural to separate the roles of the two Bobs.  In the context of summoning
quantum money, and in Examples $1$ and $2$ below of S-money schemes, we will think of the classical inputs to 
the summoning task as given to Alice by ``nature'', or by some other agency
or agencies not necessarily associated with the quantum money issuer (Bob, the Bank).   
The role of the issuer is restricted to participating in the token or money
scheme and verifying that a previously issued token is validly
presented according to the scheme rules.  
Alice wishes to satisfy the summoning task not because Bob requires
her to -- he may not care where she redeems her money, so long as it
is valid -- but because she knows, given the classical information
she received in the summoning task, that it is most advantageous to
her to redeem the money token at a particular point.  

For example, the classical inputs in the summoning task may be local prices of some
commodity, or some more general function of local market data, or
local temperatures or wind speeds, at the given space-time points.   The summoning task here 
is chosen to indicate
to Alice where it is most likely to be advantageous for her to 
present her money or token in order to carry out a transaction.

We will assume that Alice does not need to prove to Bob that the 
classical inputs took the values $m_i$ that she learned at the
points $P_i$.   Indeed she may not necessarily need even to report 
these values to Bob at all, unless the form of the token scheme
uses these values as part of the definition of the token.\footnote{
Of course, an S-money token scheme requires Alice's agents to give Bob {\it some} inputs,
and if the scheme tokens can be used to solve the type of summoning
task above, guaranteeing to Bob that the token has a unique valid
presentation point,
then they must have some relation to the $m_i$.    
However, for general $Q(m_1 , \ldots , m_M)$, the $m_i$ need not be
uniquely deducible from the valid presentation point, nor necessarily
from inputs into a scheme that validates a token at that point.}

If they do, then, for example, her local agents
at $P_i$ may be required to give the classical values $m_i$ to
Bob's local agents at $P_i$.   However the token scheme allows Bob to
accept these reported values without independent verification.
If Alice misreports in this scenario, then the token will either be
valid at a less advantageous point or not valid anywhere.
Alice thus has no motive to misreport; if she does, it is her
problem, not Bob's.  

In Examples $3$ and $4$, Alice receives inputs both from nature
and from Bob, the Bank.  These jointly define her summoning task. 
In these examples we again assume that Alice does not need to prove to Bob that the 
classical inputs from nature took the values $m_i$ that she learned at the
points $P_i$. 

\subsubsection{Quantum money and deterministic summoning tasks}

Suppose then that Alice is given a quantum money token 
containing an unknown quantum state $\psi$ at a 
point $P$ in a network, and wishes to propagate the 
quantum state over the network, exploiting the power of
general quantum operations, with the goal of satisfying 
a summoning task 
that allows her to present the money at a desired space-time point
identified as a function of incoming data at earlier points.   
Suppose further that this summoning task is possible and that
she follows a deterministic algorithm that implements it.
By ``deterministic'', we mean here that the algorithm always
propagates the state to the same presentation point for a 
given set of incoming data.   

Following this algorithm, her local agent
at any network point $Q\succ P$ 
may receive a quantum state produced by the previous actions
of local agents. 
The local agent at $Q$ may also introduce one or more new quantum states, 
which may be entangled with states that may be introduced
elsewhere in the network -- for example 
parts of singlets to be used for teleportation.\footnote{We assume
here for simplicity that $P$ is in the causal past of all network points
at which the user may wish to introduce states or carry out
operations.   However, the discussion extends to the more general
case in which some network points are not in the casual future of $P$.} 
The local agent at $Q$ may then apply a general unitary $U_Q$ which 
acts on these states collectively and propagates 
outputs along secure channels $C_i$ to further network points $R_i
\succ Q$.   
The states introduced at $Q$ and unitaries $U_Q$ may depend
on local information made available to the local agent at (or before)
$Q$. 
Local agents may also apply general measurement operations.
We can either treat these either as projective measurements (with
appropriate ancillae) in the standard projection postulate 
formalism, or enlarge the Hilbert space and consider measurements
as unitary quantum operations.   
We take the latter course, since it simplifies the discussion.   
So, although the user's algorithm may involve
operations such as teleportation that, if we apply 
the projection postulate, produce random
classical data, we treat these as deterministic unitary
quantum operations. 

By assumption, Alice's local agents follow an algorithm which determines all
these local operations as a function of local classical
information inputs $m_i$ at the input points $P_i$, in 
a manner consistent with the causal structure (so the 
operations at $Q$ depend only on inputs at points $P_i$
such that $Q \succ P_i$).    
We suppose that this algorithm guarantees that 
the quantum money token state $\psi$ will always be  
recreated at a valid presentation point
$Q_j$ if there is one, and that for any given set of 
data inputs the chosen valid presentation point is always the same.
We also suppose that the algorithm gives Alice's
agent at $Q_j$ the information that she has the
token state and that $Q_j$ is a valid presentation point. 
The agent then, having reconstructed $\psi$, may return the
quantum money token.    That is, Alice's algorithm
guarantees successfully summoning the money to the
required point, which is uniquely determined by the data inputs. 

\subsubsection{Classical emulation of  summoning quantum money}

We say a classical S-money token scheme {\it emulates} a successful
quantum summoning algorithm for quantum money if, for every possible
set of inputs to the quantum algorithm, the classical scheme has the effect of producing a valid
token at the point where the quantum summoning algorithm would have generated
and presented the quantum money state.  
Thus, instead
of Bob giving Alice a quantum money state (known to Bob but not Alice)
at the start point $P$, they initiate the classical S-money token
scheme with agreed rules.\footnote{To avoid any doubt: the classical
emulation does not require Bob to give any classical description of a quantum
money state.  The operations of the quantum money summoning
algorithm are emulated, but not the quantum states on which they act.}

To do this, they may proceed as follows. At each point $Q$, $A$'s local agent communicates
to $B$'s local agent a classical description of the choice of $U_Q$ and the nature of 
any new quantum states introduced.   For example, at $Q$ they may say ``we
introduce here singlet $n$, whose other subsystem is
available for introduction at point $Q'$''. 
If they do, then at $Q'$ 
they may say ``we introduce here singlet $n$, whose other subsystem is
available for introduction at point $Q$'', but they may instead 
say they are introducing some other state, or none.  Whether their
actions are coordinated or not depends on their decision algorithm
and on the information available to them at the relevant points.    

To present a S-money token at $Z$, $A$'s local agent describes the operation $U_Z$  
and then states that she wishes to present the token.  $B$'s local
agent at $Z$ verifies
that the token is potentially valid by checking that the set of quantum operations and introduced 
states described in the past light cone of $Z$ would indeed
have produced the state $\psi$ at $Z$.    If so, the no-cloning
theorem assures $B$ that no potentially valid token can be presented at any
point space-like separated from $Z$.   The token is thus valid, and 
may be accepted.  

Any deterministic quantum money summoning algorithm may thus
be emulated by a classical S-money token scheme.   In this sense,
S-money can solve any deterministic summoning task that quantum money can.    

\subsection{Efficient S-money schemes for summoning tasks} 

We have shown that S-money can classically emulate summoning quantum money tokens.
This gives one way in which S-money can
solve a deterministic summoning task that is solvable by quantum money summoning. 
However, there are often far more efficient S-money solutions.
\vskip 10pt

{\bf Example $ \bf 1$} \qquad Consider a task in which incoming data arrives at
a set of pre-agreed input points $\{ P_i \}_{i=1}^N$ and the S-money token may be presented at 
one of a set of presentation points $\{ Q_i \}_{i=1}^2$, with the spatial locations of 
all $P_i$ and $Q_i$ lying on the surface of a sphere (for 
example, the Earth) in a given frame.  
Here the $ P_i$ are all located on the Equator, and have time
coordinate $t=0$; $Q_1$ and $Q_2$ are located at the two poles, and 
have time coordinate $t=T$, where $T$ is the time taken to send a 
light signal from a point on the Equator to one of the poles.  
(We may either assume it is possible for light speed signals to be sent
through the interior of the sphere, or that both parties agree
this is technologically infeasible, in which
case they must go around the surface.  The time $T$ is calculated
accordingly.) 
Suppose that at $t=0$ each point $P_i$ supplies data in the form of a
subset $S_i$ of the set $\{Q_1, Q_2 \}$.  That is, each $P_i$ ``chooses'' neither presentation point,
or one of them, or both of them.  

Suppose that it is pre-agreed that if any subset of $\{ P_i \}_{i=1}^N$
belonging to some pre-agreed list of subsets $\{ S_1^1 , \ldots , S_1^{N_1}  \}$ chooses 
$Q_1$, the token is potentially valid at $Q_1$. 
Similarly, it is pre-agreed that if any subset of  $\{ P_i \}_{i=1}^N$
belonging to some pre-agreed list of subsets $\{ S_2^1, \ldots ,
S_2^{N_2} \} $ of input points chooses $Q_2$, the token is 
potentially valid at $Q_2$.  
(Here each superscript labels a distinct subset, so we necessarily have 
$N_1 , N_2 \leq 2^N$ and $N_1 , N_2 < 2^N$ for non-trivial examples; we expect
$N_1 , N_2 \ll 2^N$ for typical interesting examples.) 

We can define a S-money scheme for this task as follows.
As soon as Alice's agents at $P_i$ receive the set $S_i$, they
send the set $S_i$ to Bob's agents at $P_i$.   They also 
send the information to Alice's agents at light speed to both presentation
points.  Bob's agents at $P_i$ also send the information they receive
at light speed to Bob's agents at both presentation points. 
 The token may be
presented as soon as the signals have reached the presentation points. 
To check that the token is valid and may be presented and accepted at $Q_1$, Alice and Bob's
local agents need to check:

(a) that some
valid subset $S_1 (i)$ of input points have included $Q_1$, so that
the token is potentially valid at $Q_1$;

(b) that no valid
subset $S_2 (j)$ of input points have included $Q_2$, so that the 
token cannot also be potentially valid at $Q_2$.    

Validity at $Q_2$ 
is similarly determined.  
\vskip 10pt

\begin{figure}[h]
\includegraphics[width=0.48\textwidth]{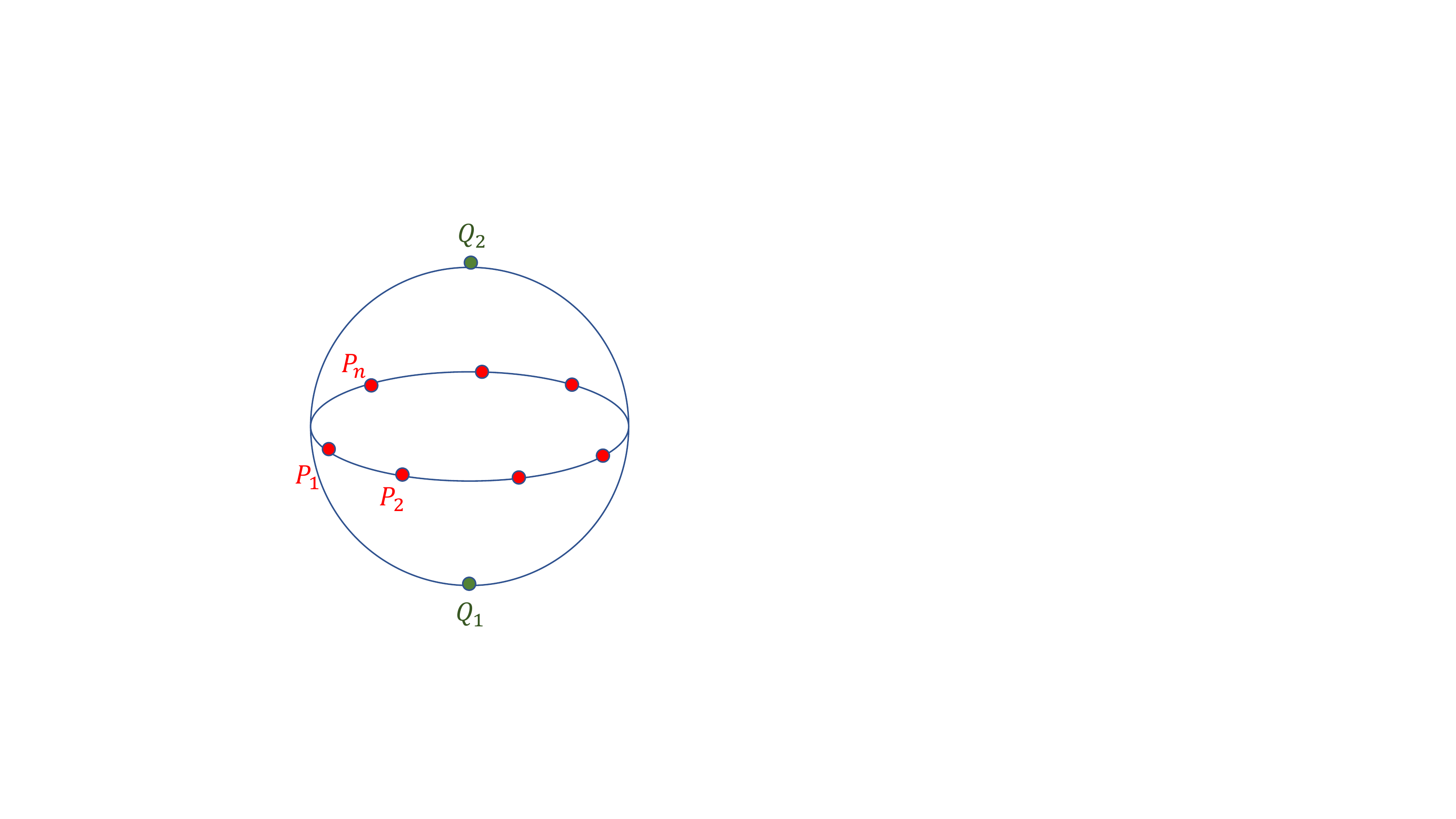} 
\includegraphics[width=0.48\textwidth]{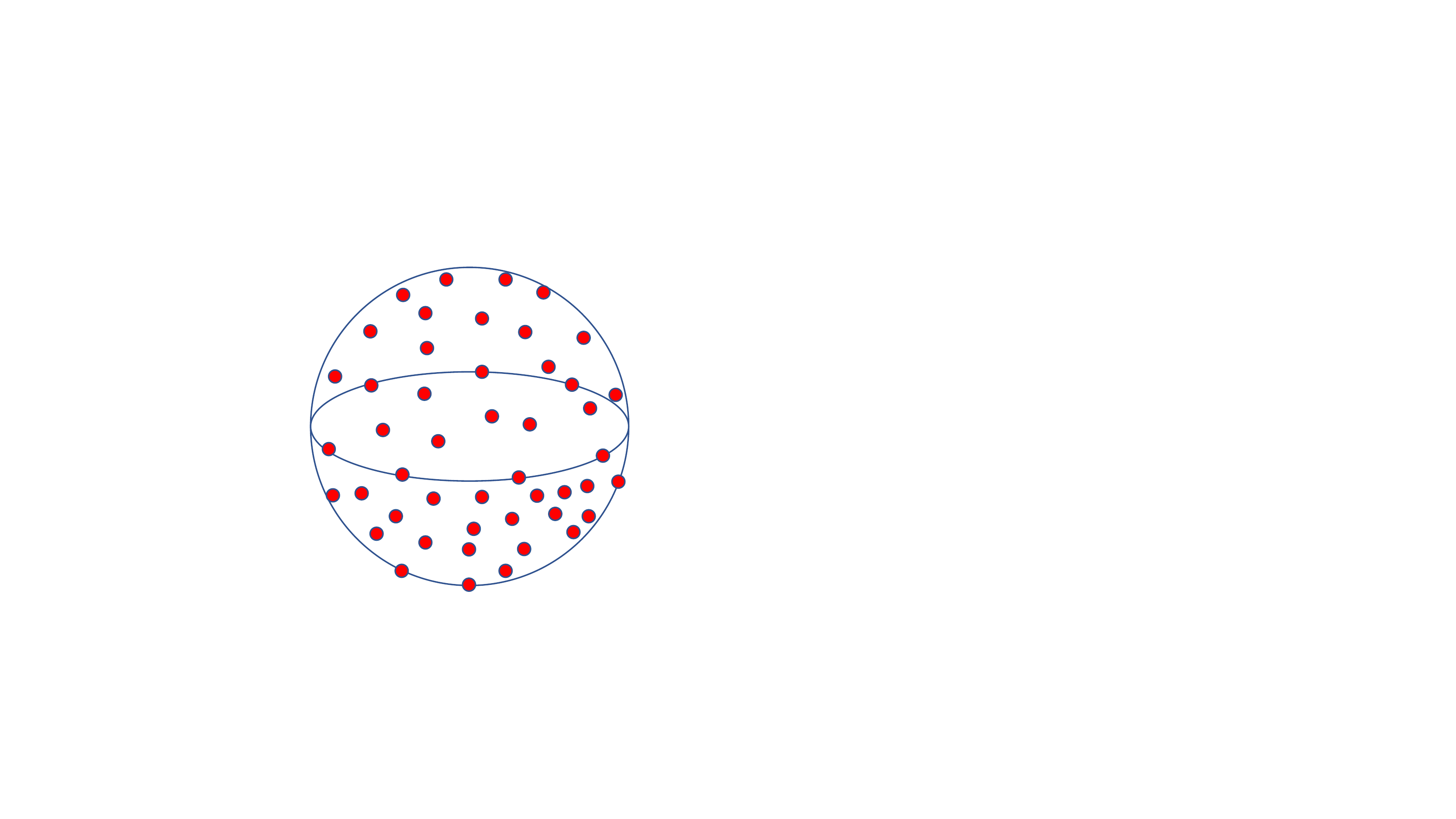} 
\caption{Left: input and presentation points in Example 1.  The spatial
locations of the input points are on the Equator, while those
of the presentation points are the Poles.
Right: input and presentation points in Example 2.  The spatial
locations of the input points are densely distributed around
the sphere.  The presentation points have the same spatial locations.}
\end{figure}

{\bf Example $\bf 2$} \qquad Consider input points whose spatial
locations are 
distributed on a spatial network
over the surface of a sphere.  We may take this network to be 
relatively dense.  For instance, if the sphere is radius $1$ 
we could suppose that every circle on the surface of radius $\epsilon$
contains at least one network point, for some $\epsilon \ll 1$.  

In this example, the spatial location of every input point is
also the spatial location of a presentation point, and 
the time coordinate of the input points is $t=0$ in an agreed frame.  
Thus the input points take the form $P_i = (\underline{x_i}, 0 )$, 
where the $3$-vector $\underline{x_i}$ represents the $i$-th spatial
location.    The presentation points are $Q_i = (\underline{x_i}, T)$, 
where $T$ is the time required for a light signal to go
between antipodal points.   (Again, the parties may agree that
communication at light speed by shortest path is possible, or
alternatively that no communication through the interior of the
sphere is possible; $T$ is calculated accordingly.)

The data presented to 
Alice's agents at each input point $P_i$ is the
value of some function $ f(P_i )$ calculated at that point. 
This function might, for example, be the price of
a stock at the relevant node of the financial network at $t=0$. 
It might also be some data generated from market data available at
$P_i$ by a
complicated algorithm that Alice does not necessarily want to
reveal to Bob.  Alice's local agent at each $P_i$ gives
the value $f( P_i )$ to Bob's co-located agent, who broadcasts them to all
Bob's agents located at the presentation points.   Alice's local agents
also broadcast the values $f( P_i )$ to all Alice's agents at the presentation
points.
At time $T$, Alice's and Bob's agents at each presentation point will have received all
the data from all the input points.   The 
token is potentially valid at a presentation point $Q_j = (\underline{x_j}, T)$, 
if the data show that
the function $f$ takes its maximum value at $P_j = (\underline{x_j}, 0)$.  The token is 
valid at $Q_j$ if it is not also potentially valid at any other 
presentation point, i.e. if the data show that the maximum at $P_j$ is unique.
(Thus, in this scheme, if there are 
$N>1$ global maxima then there is no valid presentation point for the
token.)

\subsection{Solutions with quantum money tokens}

Standard quantum money tokens can also solve the summoning
tasks given in the first two examples.    One method, applicable to both examples, is to begin with
the token at one input point and carry out a sequence of teleportation
operations through all the input points, implementing a non-local
quantum computation that results in a quantum state at the correct
presentation point which can be converted to the token state using the 
accumulated incoming teleportation data
\cite{vaidman2003instantaneous,buhrman2014position}.    

This requires the user to have agents at all input points
and to have entanglement resources that depend exponentially
on the number of sites, and near-perfect teleportation.
It thus may be far from an optimal solution in terms of resource
use.   It would be very interesting to understand how efficiently these tasks
and others of their type can be solved and to prove lower bounds on the resources
required.   We leave these questions for future work. 

Although this method for summoning quantum
money tokens requires more resources from the user (to the 
extent that it may not be technologically possible in the
forseeable future) than the S-money
scheme above, it is worth noting that it makes fewer demands
on the issuer.   If quantum money is used, the issuer's agents need
not receive inputs from the user agents during the scheme: they
need only to be ready to validate the quantum money token when
presented.  

\subsection{Solving constrained summoning tasks with S-money}

{\bf Example $\bf 3$} \qquad Consider a variation of Example $2$, with
the same spatial locations for the network points, but different S-money token rules.   In this version, 
at $t=0$, the protocol requires Bob's agents to make coordinated announcements to Alice's
co-located agents at each input point $P_i$, each identifying the same subset
$S \subset \{ P_i \}$ that defines the subset of Bob's agents that
will participate in the scheme.   (Bob might, for example, know
in advance that some of his agents will be out of action at any
given time, but not wish to reveal which ahead of time.) 
Thus, both Bob-the-bank and ``nature'' provide input data for Alice's
summoning task.   

In this variation, the S-money
token may become potentially valid after time $T'$, where $T'$ is the maximum time 
required for a light signal to travel between a pair $(P_i , P_j )$
where $\{P_i , P_j \} \subset S$.   It is potentially valid
at $P_i \in S$ if \mbox{$f(P_i ) = \max \{ f(P_j ) : P_j \in S \}$.}
It is valid at $P_i \in S$ if there is no other presentation point
at which it is potentially valid, i.e. if $P_i$ is its unique
maximum in $S$.  
(Again, this scheme allows no valid presentation point if there
are $N>1$ global maxima in $S$.) 

The S-money scheme guarantees that, if Bob behaves honestly by
declaring the same subset $S$ at each input point, then both Alice and Bob can
verify that the S-money token is valid at $T'$ at the unique maximum
point $P_i \in S$, when there is a unique maximum.   
Alice can check there that Bob's agents at every $P_i \in S$
identified the same set $S$ of participants; Bob can check
there that it is indeed the unique maximum of $f$ in $S$.  

The no-summoning theorem \cite{kent2013no} shows that there
are configurations of input and presentation points for which Alice
cannot generally solve the
task defining this scheme by summoning a quantum money token.   
For example, suppose there are two disjoint subsets $S$, $S'$ of 
$\{ P_i \}$ such that the corresponding
sets of presentation points $R = \{ Q_i : P_i \in S \}$ and
$R' = \{ Q_i : P_i \in S' \}$ have the properties that $R$ is space-like
separated from $S'$ and $R'$ from $S$. 
Suppose that Bob's agents at points $P_i \in S$
all identify $S$, while Bob's agents at points $P_i \in S'$
all identify $S'$.   
Then Alice would be required to propagate the quantum token
state to two space-like separated points, which is impossible.
Now, her ability to solve the task if Bob identifies $S$ 
at all points in $S$ is independent of what Bob's agents at points in $S'$
say, and vice versa.   Hence she cannot guarantee both to 
solve the task if all Bob's agents identify $S$ and also to do
so if they all identify $S'$. 

{\bf Comments on Example $\bf  3$} \qquad  Bob's agents
may cheat in this scheme by sending inconsistent 
information to Alice's co-located agents.   If they do, 
there may be no point in space-time at which Alice is
persuaded the S-money token is even potentially valid, in which case
she does not present or spend the token.   

Another possibility is that two or more of Alice's may be persuaded the
token is actually valid at two or more space-like separated points.
This would be the case, for example, if there are 
disjoint sets $S$ and $S'$ such that the corresponding
sets of presentation points $R = \{ Q_i : P_i \in S \}$ and
$R' = \{ Q_i : P_i \in S' \}$ have the property that 
$R$ is spacelike separated from $S'$ and $R'$ is spacelike
separated from $S$, and Bob's agents at points $P_i \in S$
all identify $S$, while Bob's agents at points $P_i \in S'$
all identify $S'$.   If so, Alice may present and try to spend
the token at two or more points; Bob then either has to 
allow the token to be spent multiply or acknowledge his
cheating at one or more of the relevant points. 

While the possibility of presenting inconsistent information in this scheme may be advantageous
to Bob in some scenarios, Alice retains the decision whether
to actually spend the token.   Also, a bank that
cheats by making inconsistent declarations 
is likely to suffer reputationally.  Alice may thus be 
willing to participate in a scheme of this type, despite 
the possibility that Bob may cheat, since 
she may see the risks and drawbacks as acceptably limited.  

An interesting variant of this task prevents Bob from sending
inconsistent inputs while preventing Alice from learning
his inputs before the time at which she may present a valid token. 
This is done  
by requiring Bob's agent at some pre-agreed point $P_0 \in \{ P_i \}$ 
to make a cryptographic commitment of Bob's input to Alice's
co-located agent, using a secure bit
string commitment scheme. 
In this variant, the token is valid
after time $T$ (where $T$ is defined as in Example $2$), so that
Alice's agent at $P_0$ may broadcast the commitment data received from
Bob's agent there to Alice's agents at all the presentation points at
$Q_i$.  Similarly, Bob's agent at $P_0$ broadcasts all the data he
received or retained from the commitment to all Bob's agents at the
$Q_i$.   Bob's agents at each presentation point $Q_i$ may then
simultaneously unveil the commitment, and Alice's located agents
may each instantaneously verify the validity of the unveiling.
The encrypted commitment prevents Alice from using the information
in Bob's input until it is unveiled, and means that the task cannot
be solved by summoning a quantum money token. 
(This remains true even if the presentation points
$Q_i$ are at time later than $T$.)
We discuss this further in Example $4$ below.

\subsection{S-money with variable value tokens and
  sub-tokens}

\subsubsection{Variable value tokens}

S-money may be extended by including rules that 
make the value of a valid S-money token depend on the information
defining the token.    For instance, in Example $2$ above,
the token value could depend on some function of the 
global distribution of the values of the function 
that are reported at the presentation point and that
validate the token there.   One simple possibility 
would be for the token
value to be the difference between the global
maximum and global minimum.    

\subsubsection{Variable value subtokens with specified validity constraints}

A further interesting class of extensions allows the
scheme to define one or more valid sub-tokens that 
may be presented at different (perhaps space-like
separated) points, and whose values may also depend
on the scheme data communicated at or to those 
points.   In this case, the issuer defines 
the security of the scheme by a set of constraints
on the values and locations of the sub-tokens. 
The scheme data communicated at or to any 
sub-token presentation point must guarantee
that no set of sub-tokens may be presented 
at locations that violate these constraints.

Such constraints may be quite general: a sub-token 
need not have value given by a fixed fraction of a token.  
For example, the scheme might allow 
either $10$ credits at a single market site,
or sub-tokens to a sum value of no more 
than $7$ credits at up to three sites on 
the same continent, or sub-tokens with
an individual value of up to $3$ credits
on sites at up to $5$ distinct continents. 
  
These features are not generally replicable by any
standard quantum money token scheme.   
For instance, in Example $2$, a quantum token
may be delocalized and propagated so that it is 
reconstructed at the unique maximum point.    However, for the user and issuer
to agree on its value there requires the user to prove
to the issuer the value of a function of the global distribution of function
values chosen at the input points.   In a S-money solution, these chosen values
are communicated directly by local user agents to local issuer agents.
Absent such communications, no set of operations by the user on the quantum
token can encode this value (for general functions) in a way that allows the issuer to 
deduce it with certainty when the output state is returned to him.

\section{ Token schemes and user privacy} 

\subsection{Privacy for tokens that follow definite paths}

Token schemes may be designed to guarantee the user various types
of privacy.   For tokens that are defined to follow definite paths in space-time
we can sensibly describe these in terms of the inferences that
can or cannot be made about points or segments of the token's path.   

Suppose that such a token is issued at point $P$ and is defined
for a network with points $V = \{ P_i \}_{i=1}^N$, 
so that it may be transmitted causally between some pairs of points on the
network, given by a set \mbox{$E \subset \{ ( P_i , P_j ) : 1 \leq i \leq N , 1 \leq j \leq
N , P_i \prec P_j \}$.}    Thus a token following the scheme rules will
follow a path $ P \rightarrow P_1 \rightarrow \ldots \rightarrow P_k
$, 
where $( P , P_1) , (P_1 , P_2 ),  \ldots , (P_{k-1}, P_k ) \in E$,
and may or may not be presented at the final point $P_k$.  
We can assign it a worldline by assigning it a linear path segment between any 
successive pair of points.    
Thus any spacelike hypersurface $S$ contains is a finite subset of 
points ${\rm Poss}(S) = \{ S_i \}_{i=1}^l$ through which a valid
token's path may possibly pass.     
 
We say such a scheme offers {\it present privacy} if it
does not require the user to give away any information about 
the token's location on any spacelike hypersurface, unless
she chooses to present the token there, beyond the information
that can be inferred from the constraints on possible paths implied by
$E$.   
Thus, if the token has not been presented in the causal past of $S$, 
and if the user follows the scheme, the collective information available 
to all issuer agents at points $Q \prec S$ or $Q \in S$ implies only that the 
token may be at any $S_i \in {\rm Poss}(S)$.   

We say the scheme offers {\it future privacy} if the information
it requires the user to give at any locations $Q \prec S$ or $Q \in S$
does
not imply any information about the token's locations at points
in the causal future of $S$, beyond the constraints implied by $E$. 

We say it offers {\it past privacy} if the information it requires
the user to give in propagating the token by any valid means from $P$ to some
network point $Q \succ P$ and in presenting it there implies no information
about any intermediate locations, beyond the
constraints implied by $E$.    That is, a presentation at $Q$ implies
only that the token followed some path $P \rightarrow P_1 \rightarrow
\ldots \rightarrow P_k \rightarrow Q$, where each successive pair 
belongs to $E$.   

\subsection{Quantum money token schemes and user privacy}

Consider now a quantum money token scheme in which the user 
is given a token quantum state at a start point $P$,  
may carry out arbitrary quantum operations at network
points $P_i$, and may present the token quantum state at
some network point $Q \succ P$.    For general networks, 
this allows the user to delocalize the quantum state,
relocalizing it before or at the point where it is presented. 
The token thus need not follow a definite path.  

Consider a standard cryptographic scenario in which the user's agents are able to carry out quantum 
operations of their choice at secure sites in the 
vicinity of each network point, and are able to 
communicate securely from network point $P_i$ to network point $P_j$ 
whenever $P_j \succ P_i$.  
In this scenario, the issuer obtains no information about
the token's presentation point, prior to presentation, 
beyond that implied by signalling constraints: it may
be presented at any network point $Q \succ P$. 
The issuer also obtains no information, after 
presentation, beyond that implied by signalling constraints
and quantum theory: the token may have been propagated 
from $P$ to $Q$ by any algorithm allowed by relativistic quantum 
theory.   

In this sense, the standard cryptographic 
scenario for quantum money tokens offers the user as complete
a form of privacy as relativistic quantum theory permits.   
This includes past, present and future
privacy as defined above in the case where the user is restricted to 
propagating the token along definite paths. 

We may extend the above definitions of past, present and future
privacy to schemes, such as
quantum money token schemes, in which the 
token has a quantum state $\psi_S$ defined on any spacelike
hypersurface $S$, with the property that the
possible user actions (including actions on the
token and presentation of the token) in the future
of $S$ depend only on $\psi_S$. 

We say such a scheme offers {\it present privacy} if it
does not require the user to give away any information about 
the token state $\psi_S$ on any spacelike hypersurface $S$, unless
she chooses to present the token there, beyond the information
that can be inferred from signalling constraints.   

We say the scheme offers {\it future privacy} if the information
it requires the user to give at any locations $Q \prec S$ or $Q \in S$
does
not imply any information about the token state $\psi_{S'}$ on any hypersurfaces
$S'$ in the causal future of $S$, beyond those implied by signalling
constraints.

We say it offers {\it past privacy} if the information it requires
the user to give in propagating the token by any valid means from $P$ to some
network point $Q \succ P$ and in presenting it there implies no information
about its state $\psi_S$ on any hypersurface $S$ lying between $P$ and
$Q$, beyond those implied by signalling constraints.     

The standard cryptographic scenario for quantum money tokens
offers the user past, present and future privacy, by these
definitions. 

\subsection{User privacy for encrypted S-money schemes}

We now consider user privacy for virtual token schemes such as
S-money.  In unencrypted S-money schemes such as those
described in Examples $1$ and $2$, the virtual token is defined by
unencrypted communications -- {\it user token inputs} -- from the user's agents to the
issuer's co-located agents at input points, and the valid presentation point
is determined by these communications.   
In Example $3$, the virtual token is defined by a combination
of the user token inputs and unencrypted communications -- {\it issuer
  token inputs} -- from the
issuer's agents to the user's co-located agents, at the same input
points.   

We want to consider encrypted versions of S-money schemes such as
these, which 
allow past, present and future privacy, in appropriate senses. 
Specifically, we consider encryptions with the following features:
\begin{enumerate}
\item each user token input may be introduced, at the relevant input point,
as the committed string in a bit string commitment protocol initiated between
the user's and issuer's local agents.   
\item data obtained from the commitment phase of the protocol by the
  user and issuer's
local agents are classical,
and thus may be broadcast to all presentation points in the causal
future of the relevant input point (the {\it commitment point}). 
\item data obtained from any subsequent sustaining phases by user
and issuer agents at appropriate points are also
  classical and may similarly be broadcast. 
\item the user's local agent may unveil the committed string at any/all presentation
  points in the causal future of the commitment point, and the
  issuer's local agent may verify the unveiled commitment at any
  presentation
point where it is unveiled.
(For some types of protocol (e.g. \cite{kentrelfinite}), the verification protocol may require a pre-agreed
time interval.  If so, this interval should be short compared to typical network scales.) 
\end{enumerate} 

We say an encrypted S-money scheme of this type offers the user {\it present and future
  privacy} if the issuer obtains no information about the 
user token inputs prior to the presentation of 
the token, beyond that implied by the scheme rules that constrain
allowed inputs.   
In particular this means that, for any spacelike hypersurface $S$, if the token
has not been presented at any point $Q \prec S$ or $Q \in S$, 
the user's inputs at all such points give the issuer no 
information about the points $Q' \succ S$ at which
the token may validly be presented, beyond that implied by
signalling constraints.      

We say the scheme offers the user {\it past privacy} if the
issuer obtains no information about the user token inputs
even after presentation of the token, beyond that implied by the
scheme rules.   
This means that, if the token is presented at $Q$, the issuer 
learns only that the user gave some valid set of inputs consistent with 
presentation at $Q$: he learns no information about which set of
inputs, if the scheme rules allow more than one

\subsection{Motivations for user privacy}

Each type of privacy may be valued for a variety
of reasons.   For token types  
that could be stolen if located, present and future privacy
enhance security.   
For tokens carried by individuals, whose location
thus closely correlates with the token's, 
past, present or future token privacy are necessary for
the corresponding forms of personal locational privacy.  

Since S-money schemes assume
that user and issuer communications are authenticated, 
they protect the parties against token theft, even in the
absence of privacy.  
However, in some scenarios, the user may only have one or a few
agents, who may value their personal locational privacy. 
Even if the user is a financial institution represented by a network of 
collaborating agents operating at known secure sites, present and future
privacy of the token scheme may be highly desirable for other reasons.   
For example, if the user's 
token represents a large sum of money to be used somewhere 
on the global financial network, with the location and time 
decided by a complex trading strategy, she may wish to conceal 
the token's present and future location from other agencies, so that they
cannot exploit advance information about her possible
trades.     
She may perhaps also value past privacy, 
so as not to reveal any more information about her
trading strategy than necessary, even after a trade.   

\subsection{Privacy in other types of money}

It should be stressed at this point that a wide range 
of cryptographic money schemes with sophisticated 
privacy mechanisms already exist, 
including for example Chaum-Fiat-Naor's ECash \cite{chaum1983blind,
chaum1988untraceable} and cryptocurrencies such as
ZCash \footnote{See https://z.cash/} and Monero
\cite{kumar2017traceability}.
As noted above, one of the advantages of quantum money schemes 
\cite{wiesner1983conjugate,
molina2012optimal,pastawski2012unforgeable,gavinsky2012quantum,aaronson2012quantum} 
is that they typically guarantee past, present and future privacy.
Public quantum money schemes (e.g. \cite{aaronson2009quantum}) and
quantum coins \cite{mosca2010quantum} give further privacy, since
transactions need not involve the bank.

We certainly do not want to suggest that
these schemes are superseded by S-money, nor that S-money 
necessarily offers equivalent privacy in all respects
and all scenarios.   Nor do we claim that the 
privacy techniques we consider below are optimally
efficient for S-money or offer the greatest degree of
privacy attainable.   We wish only to make two
positive points.  First, as we have noted, relativistic scenarios raise
interesting new privacy issues that deserve attention and require new definitions.
Second, privacy adequate for many (though not necessarily
all) purposes can in principle be incorporated into S-money.   
Since S-money has some other clear advantages in some 
interesting relativistic scenarios (and in other scenarios
with trusted signalling constraints), this encourages us to think that
it deserves consideration alongside existing proposals.   
As we noted earlier, the best schemes in network
scenarios with trusted signalling constraints 
may ultimately combine concepts and techniques 
from S-money and from other proposals.  

\subsection{Encrypted S-money with present and future privacy}

Encrypted S-money schemes add an extra layer of security,
by allowing the user to communicate all their inputs in an 
effectively encrypted form, which the issuer can only read
if the user chooses to decrypt them.   The user needs to
do this only if they wish to present the token at
a valid presentation point, and the inputs become
readable by the issuer only at that point.    
In their simplest form, the user's local agent at each input
point $P_i$ introduces her user token input as the committed
string in a single (i.e. non-redundant) bit string commitment
protocol.  The protocol commits her to the token input but does
not reveal it to the issuer's local agent.   

Bit string commitments may be
made using protocols involving the short distance transmission
of classical or quantum data, with security based on 
relativistic signalling constraints \cite{kentrel,kentrelfinite,kent2011unconditionally,kent2012unconditionally,lunghi2013experimental,kent2014secure}.
Such schemes need two or more user agents communicating with 
adjacent issuer agents; many configurations of agents are
possible.   Depending on the agent configurations and the 
choice of spacetime point at which the commitment  
may be unveiled, the schemes may require multiple rounds
of communication between each pair of agents. 
The bit string commitments could also be made by schemes 
that are not information theoretically secure 
but are presently believed to be computationally secure
(e.g. \cite{pedersen1991non}) or by ``post-quantum'' bit string
commitment schemes designed also to be computationally secure against
quantum computers.   

We require protocols in which the data obtained from the commitment phase by the
  user and issuer's
local agents are classical,
and thus may be broadcast to all presentation points in the causal
future of the relevant input point (the {\it commitment point}), 
and the data obtained from any subsequent sustaining phases by user
and issuer agents at appropriate points are also
  classical and may similarly be broadcast.  
This is true of the schemes of Refs. \cite{kentrel,kentrelfinite}.
However the security of multi-round versions of these schemes
against general quantum adversaries is not proven.  
It is also true of classical schemes that are believed to be 
computationally secure against classical or quantum computers, although of course these schemes 
are not information-theoretically secure.
For definiteness, we focus our present discussion on a variant of the scheme
of Ref. \cite{kent2011unconditionally}, for which the appropriate
information theoretic 
security is provable \cite{croke2012security,lunghi2013experimental}.  

As we explain in the supplementary material, this variant actually does not
necessarily require the user to commit herself to a given
bit string at the point where the communications
defining the commitment phase take place. 
It is thus more accurately described as a protocol 
for a closely related relativistic task, which
we term {\it bit string coordination}.   
However, this is adequate, since the security criteria for encrypted S-money
also require secure bit string coordination rather than
secure bit string commitment. 

\subsection{Privacy for issuer inputs in S-money schemes} 

We have defined present and future privacy as security desiderata for
the user, without considering the issuer's privacy.   
Indeed, in many scenarios the issuer either does not input
any information during the scheme or may input only information that he is happy to be made
public.  In either case, he has no privacy
concerns. 

However, in some scenarios, when participating in S-money schemes
such as Example $3$, in which he inputs data that affect 
when and where the token may be valid, the issuer may wish to 
ensure that these data are kept temporarily private and/or that
they are made public only if and when a token is presented.   
He may ensure this by using bit string commitment or coordination schemes to commit to the 
data without immediately revealing them.  Precisely what form of 
security the user may demand from the issuer -- bit string commitment or 
bit string coordination -- depends on the scenario.   The parties also
need to agree how to coordinate the unveilings of their respective
bit string commitments (for example by agreeing that one of them should
unveil first, or by agreeing to unveil at adjacent spacelike separated
locations) in protocols of this type.    
\vskip 10pt

{\bf Example $\bf 4$} \qquad Consider the following variation of Example
$3$, with the same network as Example $2$ but different S-money token rules. 
In this version, 
at $t=0$, Bob's agent at some pre-agreed point $P \in \{ P_i \}$ makes a commitment to Alice's
co-located agent, using a secure bit
string commitment or coordination scheme that guarantees
coordinated unveilings at all the presentation points $Q_j$.
(Recall that these have the same spatial locations as the input points,
and time coordinate $T$, the time required for a signal to travel
between antipodal points.)  
Bob's commitment identifies a subset
$S \subset \{ P_i \}$ that defines the subset of Bob's agents that
will participate in the scheme.   
Alice's local agent at each $P_i$ make coordinated commitments
of the value $f( P_i )$ to Bob's co-located agents, using a 
 secure bit
string commitment or coordination scheme that guarantees
coordinated unveilings at all the presentation points $Q_j$.

At time $T$, Bob's agents at each presentation point unveil the 
bit string commitment
identifying the subset $S$ of input/presentation points on which the
token is defined.   Alice's agents unveil their coordinated
commitments
of the values of $f( P_i )$ for the input points belonging to that subset. 

In this variation, the S-money
token is potentially valid after time $T$ at a location
$P_i \in S$ if \mbox{$f(P_i ) = \max \{ f(P_j ) : P_j \in S \}$.} 
It is valid at $P_i$ if there is no other point at which it
could be potentially valid, i.e. if $P_i$ is the unique maximum
point in $S$. 
(Thus, again, this scheme allows no valid presentation point if there
are $N>1$ global maxima in $S$.)

The S-money scheme guarantees that both Alice and Bob can
verify that the S-money token is valid at $T$ at the unique maximum
point $P_i \in S$, when there is a unique maximum.   
Bob can check
there that it is indeed the unique maximum of $f$ in $S$.  
Bob's bit string commitment or coordination scheme guarantees 
to Alice that Bob's agents at all of the presentation points $Q_j$
identify the same set $S$ of participants; it guarantees
to Bob that Alice cannot identify this set $S$ until he
unveils at the presentation points.    
Alice's bit string commitment or coordination scheme guarantees 
to Bob that Alice's agents at each presentation point $Q_j$ unveil the
same set of values $f(P_i )$ of the function at the input points. 

The no-summoning theorem \cite{kent2013no} shows that Alice
cannot generally solve the
task defining this scheme by summoning a quantum money token.   
If $S$ and $S'$ are space-like separated sets of presentation points,
she may be required to present the quantum token either at 
a presentation point in $S$ or at a presentation point in $S'$, and 
has no advance information as to which.   

\subsection{Encrypted S-money and past privacy}

In the embodiments discussed above, $A$ presents to $B$ at her final chosen
point $Q$ a classical string of bits defining commitment data for
the various commitments that encrypt the data defining the S-money token. 
$B$ can use these commitment data, together with his validation data,
to unveil the commitments, and thus obtain the unencrypted S-money
token input data.
This allows $B$ to validate the S-money token at $Q$, but also gives $B$ 
all the token input data.  
 
To prevent this and ensure past privacy, $A$ and $B$ may proceed
as follows.   $B$ may present to $A$ the requirements for a token to be valid
at $Q$ in the form of a testing algorithm, which may for example include
statistical tests that allow for some level of errors and for
statistical variance in the outcomes of any quantum measurements
$A$ may have made in the course of the commitment protocols. 
This algorithm may conveniently be agreed in advance, although 
in principle it could be specified by $B$ at $Q$. 
Instead of sending commitment data directly, $A$ may 
encrypt it using any standard 
cryptographically or unconditionally secure bit commitment scheme,
in a redundant form that allows $A$ and $B$ to follow a zero-knowledge
proof protocol (e.g. Ref. \cite{brassard1988minimum}) that simultaneously
guarantees to $B$ that the commitment data correspond to a token
valid at $Q$ and guarantees to $A$ that $B$ learns no information
about the S-money token input data other than the information
implied by its initial state, its validity at $Q$, the rules of the 
given S-money scheme, and the no-superluminal signalling principle.

\section{Security and Practicality: Issues and Extensions}

We have suggested a conception of secure S-money tokens in space-time
implemented by local communications between user and issuer agents, 
and characterised by security properties of the token scheme.    
The key feature of our schemes is a guarantee against multiple
presentation of apparently valid S-money tokens.   This is ensured
by the logical structure of the schemes.   The schemes require
communications to be completed in small regions around each
network point.   In their simplest versions, they require the issuer agents to transmit 
information received from user agents to all issuer agents
in their causal future, who must store it until it can no
longer be relevant.   Depending on the details of the token
scheme, this may require storing it until they learn that 
the token has been validly presented.   
These communication and storage requirements imply 
technology-dependent bounds on the network density and on token
lifetimes.   However, the schemes can clearly be implemented
securely for many interesting and realistic examples of networks
and token lifetimes.

Extensions of the scheme to allow present, future and past
privacy raise new questions.   
Unconditionally secure bit commitment schemes
can be proven secure, against both classical and quantum adversaries,
assuming only the validity of quantum theory and special relativity, when used to commit a single bit, 
at least for a single round \cite{kentrelfinite,kent2011unconditionally,kent2012unconditionally,lunghi2013experimental,kent2014secure}.
These schemes are also unconditionally and composably secure 
against the receiver, Bob, since from his perspective the 
information he receives \footnote{If any -- in some of these schemes 
he receives no information until unveiling.} for each committed bit is
random, and is uncorrelated between commitments.  
Specific zero knowledge protocols based on relativistic
bit commitments have also been proved secure \cite{chailloux2017relativistic}. 
However, security and efficiency questions for general zero knowledge protocols
based on relativistic bit commitment schemes remain open.
A fortiori, this is also true
for protocols that incorporate bit commitments and zero knowledge 
proofs as subprotocols.
S-money token schemes add further motivation for 
exploring these intriguing questions.

As we have noted, secure S-money schemes need to ensure 
that the decrypts of encrypted data are securely coordinated, not necessarily
that the user was securely committed to these data at some specified input
point.    
Still, the composable security of such schemes against a committer
equipped with unlimited quantum technology, who may be unveiling 
many commitments, each at many sites, remains to be fully
analysed.
It also remains to be understood
how to optimize such protocols for efficiency, so 
that a given level or type of security can be 
guaranteed with minimal (classical and/or quantum) communication, storage,
entanglement or information 
processing resources.   

As noted earlier, S-money token schemes could alternatively
be based on classical non-relativistic bit commitment schemes that are believed to
be computationally secure against classical or quantum computers.  Composability issues are 
a much less serious concern (albeit not proven irrelevant)
if such schemes are used, and 
we believe present technology should allow our these
versions of our schemes
to be implemented in some interesting and practically
relevant examples with good security.   

While we have shown that interesting S-money schemes with significant
advantages may be based on bit commitment, bit coordination and zero
knowledge protocols, we do not claim these techniques are generally
optimal.   Our results and examples should be taken as existence
theorems that may motivate work on potentially more efficient or
alternatives.   For example, it would be interesting to explore
using appropriate secure multi-party computations to guarantee
the validity of a token at the presentation point without 
revealing the token input data.\footnote{I thank a referee for this 
suggestion.}

Our schemes have been presented in a scenario where there is a 
well-defined network with an issuer agent at every node.  
They can be extended to scenarios where issuer and user
agents may not have precisely guaranteed location, and
may both be mobile.  For example, user agents may broadcast authenticated
token input data, to be received by issuer agents wherever they 
are.   An issuer agent will accept a presented token as valid only
when (a) they have received the presentation data for the token,
(b) all the relevant token input data has been relayed to them by other
issuer agents.    Of course, this will generally involve delays
between presentation and acceptance.
Unless the user has some guarantees about the location of issuer
agents, they have no guarantee about the duration of these delays.
However if, for example, each issuer agent is guaranteed to be within a 
specified region of small radius, these delays may be guaranteed to
be small.    It would be interesting to study more general examples.

In our network model in which local user and issuer agents have 
authenticated channels, S-money tokens or sub-tokens can be
transferred from one user to another by local authenticated
communications.   Indeed, each individual token input datum
can be separately transferred, whether or not it defines a 
sub-token in the scheme.  For example, Alice's local agent at $P_i$
can tell Bob's local agent $B_i$ that her local token data is now
associated with Charlie, and $B_i$ can confirm this with
Charlie's local agent $C_i$.   $B_i$ communicates this to 
Bob's agents in the causal future of
$P_i$, who will no longer accept this input datum as part 
a valid token presentable by Alice, but will accept it as
part of a valid token presentable by Charlie. 
In general, transferring a complete S-money token thus 
requires coordinated actions between triples $(A_i , B_i , C_i$ 
at more than one network point $P_i$.    
Transferring a delocalized quantum money token 
also requires coordinated actions between pairs $(A_i , C_i )$
at more than one $P_i$.   However, an advantage of quantum
money is that the agents $B_i$ need not be involved.
This may not be a significant advantage in scenarios 
where all the relevant parties have agents at all
network points and where the parties are not concerned
about keeping ownership private from the bank, but 
there are certainly other natural scenarios in which it 
is important.
It would thus be particularly interesting to investigate 
other mechanisms for transferring S-money tokens.   

\section{Conclusions} 

We have proposed defining secure S-money token schemes that 
(i) prevent space-like 
duplication, (ii) allow the solution of generalised
summoning tasks, including tasks that cannot
be solved by standard quantum money schemes.   
We have also proposed extensions of these schemes involving general types of subtoken
that also satisfy (i-ii) and have very flexible properties. 
These token and sub-token schemes have implementations 
with current technology for many types of practically relevant network.  

We have also proposed refinements of these schemes that offer past,
present and future privacy, modulo questions about composable  
security that remain to be fully addressed.  
Their optimal implementations remain to be characterised, 
as does the full range of tasks for which they
can be practically implemented with present technology. 

Quantum money schemes offer other ways of satisfying (i) and of 
solving some generalised summoning tasks, and guarantee
past, present and future privacy to the extent quantum theory allows.  
While S-money token schemes can in principle solve any deterministic summoning
task that can be solved with quantum money, 
the converse is not true.  
Quantum money schemes with long token lifetimes are not
feasible with current technology.   

Intriguing theoretical and practical questions about the
relative advantages, implementability and resource costs of S-money
and quantum money token schemes thus remain.   We hope this will motivate further theoretical
and experimental work.   

{\bf Comment} \qquad  Refs. \cite{patents} contain some further
examples and discussions.   

{\bf Funding statement} \qquad This work was partially supported by 
UK Quantum Communications Hub grant no. EP/M013472/1 and by 
Perimeter Institute for Theoretical Physics. Research at Perimeter
Institute is supported by the Government of Canada through Industry
Canada and by the Province of Ontario through the Ministry of Research
and Innovation.

\bibliographystyle{unsrtnat}
\bibliography{relmoney.short.bib}{}

\section{Supplementary material: Bit string coordination and bit string commitment} 

In our main scenario for S-money, the user inputs data 
at various space-time points, and these inputs collectively determine the
space-time point at which the S-money token will be valid. 
Effectively, the user freely controls the destiny of the S-money
token, and the scheme is designed to give as much freedom to the user as
possible while preventing fraudulent presentation of duplicate tokens.
In this scenario, it is not an additional security requirement that the user 
necessarily must be committed to a specific destination by choices
made at specific pre-agreed points in space-time.   

So, if the user's inputs are
encrypted, the necessary security requirement for S-money is that the user should not
be able to produce different apparently valid decrypts of the same
input at different space-time points (and in particular at space-like
separated points).   This does not necessarily imply that the user
was committed to a specific decrypt value at the input point.  
It is thus possible in principle to implement secure S-money schemes, and
to maintain present and future user privacy, while using 
subprotocols that guarantee {\it bit string coordination}, meaning that 
any two apparently valid decrypts of a bit string must (with suitably high
probability) be identical.   This can be done without 
necessarily guaranteeing {\it bit string commitment} at some specified input
point. 

The weaker security requirement is significant even
if protocols that guarantee secure bit commitment, and not only secure 
bit coordination, are used in a S-money scheme. 
There are several reasons for this.
One is that the security parameters for the relevant schemes need
be chosen only so as ensure appropriately secure bit coordination
rather than bit commitment.
Another is that, for schemes with incomplete security proofs against
general attacks, we only need to prove that the scheme securely
implements bit coordination rather than bit commitment.   
Moreover, the composable security of bit commitment 
protocols in a S-money scheme need only be analysed
with respect to composably secure coordination rather than
composably secure commitment.      

\subsubsection{Definition of bit coordination}

A {\it bit coordination} scheme in space-time is a protocol 
defined in the standard scenario for mistrustful relativistic cryptography.
That is, Alice is represented by a network of collaborating and 
mutually trusting agents distributed 
throughout space-time, occupying secure sites; Bob is represented
by a similar network.   The secure sites occupied by Alice's
agents are disjoint from those occupied by Bob's.
The protocol may (and in the cases we focus on does) prescribe
a sequence of coordinated communications between adjacent agents of
Alice and Bob taking place within prescribed time windows.   However,
we assume the secure sites of adjacent agents are so close in space
and the communications so close in time, compared to network point
separations and other protocol time parameters, that we
may sensibly consider them as effectively taking place at a single space-time
point.  

At some finite set of initial points $\{ P_i \}_{i=1}^m$, Alice's
and Bob's co-located agents exchange classical and/or 
quantum information, following a prescribed protocol, 
which may require either or both to make random choices
from prescribed distributions.   

At another set of final points $\{Q_j \}_{j=1}^n$, Alice's 
and Bob's co-located agents make further exchanges
of classical and/or quantum information, again
following a prescribed protocol.  These include
a declaration of a bit value $b_j \in \{ 0 , 1 \}$ 
by each of Alice's agents.   At each $Q_j$, 
the protocol stipulates an algorithm in which $B_j$
inputs all the information he received or retained as part 
of the protocol, including the value $b_j$, and
outputs $O_j$ taking the value ${\bf accept}$ or ${\bf reject}$.  
If there are no constraints on Bob's signalling other than those
implied by the causal structure of space-time, we may take the final
points $Q_j$ to be mutually space-like separated, since
if Alice declares a bit value $b_i$ at $Q_i$ then 
Bob's agent there may send the declared value $b_i$ and his 
output $O_i$ to any point
$Q \succ Q_i$.  

Alice's agents may carry out any classical
or quantum operations on
any information they generate or receive and may
transmit such information to other agents of Alice. 
Such operations and communications may be required
as part of the protocol and may be required to be
implemented securely; we assume secure devices and
channels are available as required.   
The same holds true for Bob. 
The protocol stipulates non-empty sets $S_b = \{ A^i_b : i \in I \}$ of coordinated 
sets of actions, $A_b^i$, for Alice's agents, where the same indexing
set
$I$ labels $S_0$ and $S_1$.  Alice may choose any
set of actions $A^i_b \in S_b$ if she
wishes all her agents to unveil bit value $b$.   Each $A^i_b$ includes
the prescription that each $A_j$ should declare bit value $b$ at
$Q_j$.   

We say the protocol is {\it  $\epsilon$-sound} if, 
when Bob honestly follows the 
protocol and Alice follows any $A_b^i \in S_b$, 
the probability that all of Bob's agents accept
obeys
\begin{equation}
{\rm Prob} (O_1 = \ldots = O_n = {\rm \bf~accept~}) \geq 1 - \epsilon \,
. 
\end{equation}

We say the protocol is {\it $\epsilon$-secure} against
Alice if, whatever strategy she uses, including strategies not in
$S_0$ or $S_1$, 
\begin{equation}
{\rm Prob} ( \exists j \neq j' {\rm~such~that~} b_j \neq b_{j'} \, \,
\& \, \, O_j
= O_{j'} = {\rm \bf~accept~} ) \leq \epsilon \, .
\end{equation}

We say the protocol is {\it $\epsilon$-concealing} if
the following condition holds.   
Suppose that Alice follows
prescription $A^i_b$ for some given $i \in I$ known to Bob
and a bit value $b \in \{0, 1 \}$ of her choice,
everywhere up to the unveiling points $Q_j$, where Alice's
agents do not send any information.  
Suppose that Bob's agents then share all
the information received during the protocol,
communicating it all to a single agent $B'$ at some point $Q'$,
where $Q' \succeq Q_J$ for all $j$.   
Then $B'$ can learn no more than $\epsilon$ bits of information
about Alice's chosen bit $b$.  

Note that, while the protocol stipulates a set $S_b$ of
actions for Alice's agents to carry out to ensure that
they can almost surely each declare bit value $b$ 
and have their declarations accepted, this set
need not be an exhaustive list.
An $\epsilon$-sound and
$\epsilon'$-concealing protocol may also allow Alice
other strategies in which she chooses a bit value $b$,
which her agents all unveil, such that all Bob's agents
accept the unveiling with probability at least $(1 - \epsilon )$
and such that Bob can learn no more than $\epsilon'$ bits of
information about the chosen bit if it is not unveiled.  

Note too that Alice may have strategies that almost
surely result in her agents' all unveiling the same bit
and Bob's agents all accepting these unveilings, even
though some or all of her agents do not know the value
of the bit to be unveiled in advance.   She may, for
example, prepare a state of the form 
\begin{equation}
\sum_{b=0,1} a_b \ket{b}_{A_1} \ldots
\ket{b}_{A_n}  \ket{b}_{I_1} \ldots \ket{b}_{I_N} \, ,
\end{equation}
where $\sum_{b=0,1} | a_b |^2 =1$, $a_0 \neq 0 \neq a_1$, the state $\ket{~}_{A_i}$ is an 
ancilla state sent to Alice's agent at $Q_i$ and the $\ket{~}_{I_j}$ states are input into
the protocol as required, where input $\ket{b}$ 
corresponds to Alice choosing bit value $b$. 
Alice's agents at $Q_i$ may each measure their $\ket{~}_{A_i}$ state, 
each obtaining the same bit value $b$, which they declare along with
the unveiling data prescribed by the protocol.   This superposition
strategy is $\epsilon$-sound, $\epsilon'$-secure and
$\epsilon''$-concealing
if each of its components is.
None of Alice's agents know the value of $b$ until the point $Q_i$. 
Nonetheless, this does not {\it per se} violate the security criteria.
A secure protocol guarantees to Bob that he will almost certainly reject
if any two of Alice's agents unveil different bit values; it 
is not required to guarantee to him that the agents knew these values in 
advance of unveiling.    

\subsubsection{Definition of bit string coordination}

We define a {\it bit string coordination} scheme in space-time
similarly.   As before, at some finite set of initial points $\{ P_i \}_{i=1}^m$, Alice's
and Bob's co-located agents exchange classical and/or 
quantum information, following a prescribed protocol, 
which may require either or both to make random choices
from prescribed distributions.   

At another set of final points $\{Q_j \}_{j=1}^n$, Alice's 
and Bob's co-located agents make further exchanges
of classical and/or quantum information, again
following a prescribed protocol.  These include
a declaration of a bit string value 
$$s_j = b_{j,0} \ldots b_{j,{t-1}} \in \{
0 , 1 \}^t$$
by each of Alice's agents.   At each $Q_j$, 
the protocol stipulates an algorithm in which $B_j$
inputs all the information he received or retained as part 
of the protocol, including the value $s_j$, and
outputs $O_j$ taking the value ${\bf accept}$ or ${\bf reject}$.  
Again, if there are no constraints on Bob's signalling other than those
implied by the causal structure of space-time, we may take
the final points to be mutually space-like separated. 

As before, Alice's agents may carry out any classical
or quantum operations on
any information they generate or receive and may
transmit such information to other agents of Alice. 
Such operations and communications may be required
as part of the protocol and may be required to be
implemented securely; we assume secure devices and
channels are available as required.   
The same holds true for Bob. 
The protocol stipulates non-empty sets $S_{s} = \{ A^i_s : i \in I \}$ of coordinated 
sets of actions, $A_s^i$, for Alice's agents, where the same indexing
set
$I$ labels each $S_{s}$.  Alice may choose any $A^i_s \in S_s$ if she
wishes all her agents to unveil string value $s$.   Each $A^i_s$ includes
the prescription that each $A_j$ should declare string value $s$ at
$Q_j$. 

We say the protocol is {\it  $\epsilon$-sound} if, 
when Bob honestly follows the 
protocol and Alice follows any $A_s^i \in S_s$, 
the probability that all of Bob's agents accept
obeys
\begin{equation}
{\rm Prob} (O_1 = \ldots = 0_n = {\rm \bf~accept~}) \geq 1 - \epsilon \,
. 
\end{equation}

We say the protocol is {\it $\epsilon$-secure} against
Alice if, whatever strategy she uses, including strategies not in any
$S_s$, 
\begin{equation}
{\rm Prob} ( \exists j \neq j' {\rm~such~that~} s_j \neq s_{j'} \, \,
\& \, \, O_j
= O_{j'} = {\rm \bf~accept~} ) \leq \epsilon \, .
\end{equation}

We say the protocol is {\it $\epsilon$-concealing} if
the following condition holds.   
Suppose it is given that Alice will follow
prescription $A^i_s$ for some given $i \in I$ known to Bob
and a string value $s \in \{0, 1 \}^t$ of her choice,
everywhere up to the unveiling points $Q_j$, where Alice's
agents do not send any information.  
Suppose that Bob's agents then share all
the information received during the protocol,
communicating it all to a single agent $B'$ at some point $Q'$, 
where $Q' \succeq Q_J$ for all $j$.   
Then $B'$ can learn no more than $\epsilon$ bits of information
about Alice's chosen string $s$.  

As in the case of bit coordination, the protocol may allow Alice
other sound and concealing strategies, including
strategies in which her agents coordinate to reveal
a string that remains in a superposition state until
unveiling. 

\subsubsection{A secure bit string coordination protocol}

To define a secure bit coordination protocol, consider the following variant of the scheme of
Ref. \cite{kent2012unconditionally}.   A secure bit string
coordination protocol may be defined simply by composing a number
of these secure bit coordination protocols.   
Alice and Bob agree on a space-time point $P$
and $n$ space-like separated points $\{ Q_i \}_{i=1}^n$ in the causal future of $P$. 
They each have agents, separated in secure laboratories, adjacent to 
each of the points $P, Q_1 , \ldots ,  Q_n$.  To simplify for the moment,
we take the distances from the labs to the relevant points as
negligible.   Alice may have other agents at other locations.
We assume that Alice's agents have secure quantum and classical
communication channels as required.   
Alice's secure classical channels could, for example, be created
by pre-sharing one-time pads between her agent at $P$ and those
at $Q_0$ and $Q_1$ and sending pad-encrypted classical signals.
If necessary or desired, these pads could be periodically replenished by 
quantum key distribution links between the relevant agents. 
Her secure quantum channels could, for example, be teleportation
using pre-shared entangled states and (not necessarily secure, but
authenticated) classical communication. 
We assume that Bob's agents either have secure
classical communication channels or pre-share the classical
description of the qubit string used by Bob in the protocol.  

Bob securely preprepares a set of qubits $\ket{\psi_i}_{i=1}^N$ independently
randomly chosen from the BB84 states $\{ \ket{0} , \ket{1} , \ket{+},
\ket{-} \}$ (where $\ket{\pm} = \frac{1}{\sqrt{2}} ( \ket{0} \pm
\ket{1} )$) and sends them to Alice to arrive (essentially) at $P$. 

We define a strategy $S_{P'}$ for Alice 
to produce a coordinated unveiling of the bit value of her choice, 
for any point $P'$ such that $P \preceq P' \preceq Q_i$ for all $i$.  
To be able to unveil $0$ in strategy $S_{P'}$, an agent of
Alice's agent at $P'$ measures each state in the $\{ \ket{0} ,
\ket{1} \}$ basis, and sends the outcomes over secure classical 
channels to her agents at each of the $Q_i$. 
To be able to unveil $1$, Alice's agent at $P'$ measures each state in the $\{ \ket{+} ,
\ket{-} \}$ basis, and sends the outcomes as above. 

To unveil her coordinated bit, Alice's agents at each of the $Q_i$
reveal her measurement choices and outcomes to Bob's agents there. 
Each of Bob's agents checks that the choices and outcomes revealed to them are statistically consistent with the 
qubit string initially sent and that the choices correspond to the
declared bit value.    If so, they accept the revelation; if
not, they reject. 

For Alice to be able to successfully cheat on any coordinated bit, at 
least two of her agents must be able to produce measurement choice
and outcome data statistically consistent with measurements in
the two BB84 bases on the same set of $N$ qubits.   
There are $n(n-1)/2$ pairs of agents, and the probability
of any pair successfully doing so \cite{croke2012security}
decreases exponentially with $N$.  The protocol is thus
$\epsilon (N)$-secure against Alice, where $\epsilon(N) \rightarrow 0$
as the security parameter $N \rightarrow \infty$.   

Alice gives Bob no information until unveiling, so the 
protocol is perfectly (i.e. $0$-) concealing.  

In the ideal case of perfect preparation, transmission 
and measurement, Alice's results are always statistically
consistent with her declared bit value and measurement
choices.  The protocols is thus perfectly (i.e. $0$-) sound
in this case.   

The protocol can also be shown to be secure and sound 
in the presence of non-zero errors and losses \cite{croke2012security,lunghi2013experimental}.  

Notice that, if any pair $Q_i$ and $Q_j$ are space-like
separated, Alice's successful unveiling at $Q_i$ and $Q_j$
guarantees that she was committed to the unveiled bit,
in the standard sense \cite{kent2011unconditionally}, 
at some point in the past of $Q_i$ and $Q_j$.   
However, it does not identify the point at which she
was committed.  In particular,
it does not guarantee that she was committed at $P$.

\end{document}